\documentclass[%
 aps,
 pre,%
 amsmath,amssymb,
 reprint,%
]{revtex4-1}
\usepackage{xcolor}
\usepackage{graphicx}
\usepackage{dcolumn}
\usepackage{bm}
\usepackage[bookmarks=false,hyperfootnotes=false]{hyperref}
\hypersetup{
			colorlinks=true,
			linkcolor=blue,
			anchorcolor=black,
			citecolor=blue,
			urlcolor=blue,
}
\usepackage{multirow}
\newcommand{\dt}{\partial_{t} }
\newcommand{\dex}{\partial_{x} }
\newcommand{\dey}{\partial_{y} }
\newcommand{\dez}{\partial_{z} }
\newcommand{\ddey}{\partial^{2}_{y} }

\newcommand{\hv}{\hat v}

\newcommand{\he}{{\hat \omega_{y}}}

\newcommand{\dtot}[2]{\frac{\mathrm{d} #1}{\mathrm{d} #2}}
\newcommand{\BU}[1]{\mathop{}\!\bm{\mathrm{ #1}}}
\newcommand{\ud}{\mathop{}\!\mathrm{d}}  
\definecolor{verde}{rgb}{0.0,0.7,0.0}

\begin{document}

\preprint{AIP/123-QED}

\title{Dispersive to \textcolor{black}{non-dispersive} transition and phase velocity transient for linear waves in plane wake and channel flows.} 



\author{Francesca De Santi$^1$, Federico Fraternale$^1$, Daniela Tordella$^1$}
\thanks{\textcolor{blue}{Accepted for publication on Physical Review E} \\Email address for correspondence: daniela.tordella@polito.it}
\affiliation{$^1$ Department of Mechanical and Aerospace Engineering, Politecnico di Torino,
Torino, Italy 10129}


\date{\today}

\begin{abstract}
\textcolor{black}{In this study we analyze the phase and group velocity of three-dimensional linear traveling waves in two sheared flows, the plane channel and the wake flows. This was carried out by varying the wave number over a large interval of values at a given Reynolds number inside the ranges $20-100$, $1000-8000$, for the wake and channel flow, respectively. Evidence is given about the possible presence of both dispersive and \textcolor{black}{non-dispersive} effects which are associated with the long and short ranges of  wavelength.
We solved the Orr-Sommerfeld and Squire eigenvalue problem and observed the least stable mode. 
It is evident that, at low wave numbers, the least stable eigenmodes in the left branch of the spectrum beave in a dispersive manner. By contrast, if the wavenumber is above a specific threshold, a sharp dispersive to \textcolor{black}{non-dispersive} transition can be observed. Beyond this transition, the dominant mode belongs to the right branch of the spectrum. 
The transient behavior of the phase velocity of small three-dimensional traveling waves was also considered. Having chosen the initial conditions, we then show that the shape of the transient highly depends on the transition wavelength threshold value. We show that the phase velocty can oscillate with a frequency which is equal to the frequency width of the eigenvalue spectrum. Furthermore, evidence of intermediate self-similarity is given for the perturbation field.}

\end{abstract}

\maketitle

\section{Introduction}

The relationship between the oscillation frequency and the wave vector - which is known as the dispersion relation - is a general relationship that is important in all areas of wave physics.  Simply stated, a dispersion relation is the function $\omega(k)$ for a harmonic wave. For the most basic waves, where the speed of propagation $c$ is a constant, the dispersion relation is simply $\omega(k) = c k$. That is, the angular frequency is a linear function of the wave vector. The ratio $\omega/k$ is the propagation speed c and it is known as the phase velocity. In this case, the phase speed is independent of $k$. 
However, this is not always the case. An example is given by the propagation of light in a dielectric medium, where the index of refraction $n = c_0/c$ (where  $c_0$ is the speed of light in a vacuum) depends upon the wavelength. Another case of an interesting, nonlinear dispersion relation is found in the wave function that describes a free particle propagating in a given direction with a given momentum.

In the context of temporal linear perturbation dynamics of viscous incompressible flows, the general solution to the Orr-Sommerfeld equation at a fixed Reynolds number yields a dispersion relation between a real wavenumber $k$ and a complex angular frequency $\omega$.  Any arbitrary initial disturbance can be decomposed and expressed as a linear combination of eigenfunctions. For bounded flows, the Orr-Sommerfeld spectrum consists of an infinite number of discrete eigenvalues and it is complete \cite{dh1969}. While for unbounded flows, only a finite number of discrete eigenvalues  exists and thus a continuum of modes must exist \citep{gs1978, sg1981}. 

In the long term the least stable mode (LSM), i.e. the mode which has the maximum temporal growth rate,  prevails. In temporal stability analysis, the focus is usually placed on the range of wavenumbers where the maximum temporal growth rate $\omega_i$  is positive and in general  the information on the relation $\omega_i =\omega_i(k)$ is more abundant with respect to the corresponding $\omega_r =\omega_r(k)$ \cite{gs1968, gs1978, g1991, craik1985, cjj2003, sh2001}.

\textcolor{black}{Spectral theory of hydrodynamic waves has been developed to a high degree of sophistication, with intricate analytical tools supplemented by accurate multi-dimensional computational methods \cite{t2005,mt2003,ttrd1993,rh1993,hr1994}.
Nevertheless, unresolved key issues in this classical problem can be still identified. For instance, let us consider the following questions:
%
%
\newline -- Perturbations at a fixed wavelength and Reynolds number don't always retain the same frequency of oscillation during their transient. Thus the phase speed may vary inside transients. How are these variations linked to the structure of the eigenvalue spectra, in particular to their branched structure?
\newline -- 
In open sheared flows, like wakes for instance, at fixed values of the control parameter, differences in the dispersiveness associated to the set of eigenvalues belonging to the left and right branches of the spectrum are observed. A fact which has not been yet examined in the case of wall flows. In synthesis, what happens to the least stable mode (LSM) when the perturbation wavelength is varied by fine steps over a large range? Does the \textcolor{black}{dispersiveness} change when the LSM moves from one branch to the other?  
\newline -- Is intermediate self-similarity of the perturbation profiles present inside transients?
\vspace{2mm}
\newline Specifically, we have considered these issues to take aim at two points.}
%
Firstly, the relationship between the angular frequency corresponding to the least stable Orr-Sommerfeld mode and the perturbation wavenumber was analyzed, in a wide range of wavenumbers, that is from $k =0.2 0$ to $k=20$. This interval extends from below to well above the eventual instability range.  We consider two classical incompressible shear flows, the plane channel and wake flows. The analysis was carried out by considering Reynolds numbers in the range $1000-8000$ for the channel flows, and in the range $20-100$ for the wake. The aim was to obtain both the phase speed and the group velocity and by comparison observe possible variations of the dispersive behavior with the wavenumber.

We observed the existence of a wavenumber threshold ($k_d$, in the following) that separated the waves which propagate in a dispersive way from the waves that propagate in a non-dispersive way. 
While of interest for the study of wave packets, the existence of this \textcolor{black}{dispersive/non-dispersive} transition can also affect the transient dynamics of the traveling waves. 

Secondly, the temporal evolution of sheared flow small perturbations is investigated, with a particular emphasis on the phase velocity transients and self-similarity issues. Indeed, the frequency or phase velocity transient has been poorly investigated to date. For instance, in the wake flow, attention has mainly focused on the frequency of vortex shedding for the most unstable spatial scales \cite{w89,p1999,ss1990}. Only very recently, subcritical wake regimes (up to values $30\%$ below the critical value) of transiently amplified perturbations have been studied by considering the spatio-temporal evolution of wave packets \cite{m2011}. The situation is quite different within the context of atmosphere and climate dynamics. Here, the interaction between low-frequency and high-frequency phenomena, which is related to the existence of very different spatial and temporal scales, is believed to be one of the main reasons for planetary-scale instabilities \cite{s2002,na1997}. However, due to the inherent strong nonlinearity, the evolution of single scales cannot be observed in geophysical systems and thus also these studies usually do not account for the structure of the frequency transient of a single wave. 
In this work, the temporal evolution between the early transient and the \textcolor{black}{temporal asymptotic} state was considered in detail. It is observed that this transition can be characterized by phase velocity jumps inside the perturbation temporal evolution. This implies that the perturbation may experience acceleration or deceleration. This behavior depends on the initial condition adopted and on the wavelength of the perturbation, in particular if this is smaller or greater with respect to the \textcolor{black}{dispersive/non-dispersive} threshold.

The organization of the paper is as follows. In section 2 both an eigenvalue problem and an initial-value problem are formulated, (see also Appendix A). The dispersion relation \textcolor{black}{of the least stable mode}  is discussed in Section 3. The phase velocity behavior in the transient is then described in Section 4. In Section 5, a discussion on the existence of the intermediate term and its empirical determination is presented. Conclusion remarks follow in Section 6.  

\section{Physical problem and formulation}
We consider two typical shear flows, the plane channel flow, an archetype of bounded flows, and the plane bluff-body wake, one of the few free flow archetypes. A Cartesian coordinate system is adopted, with origin at the channel mid plane in the first case and at the bluff-body location, for the wake case. The  $x,y,z$ axis are oriented in the streamwise, the transversal and the spanwise directions, respectively  (see Fig. \ref{Fig_1}). After introducing arbitrary small  perturbations the linearized, viscous and incompressible governing equations in non-dimensional form read:
\begin{figure}
\includegraphics[width=0.8\columnwidth]{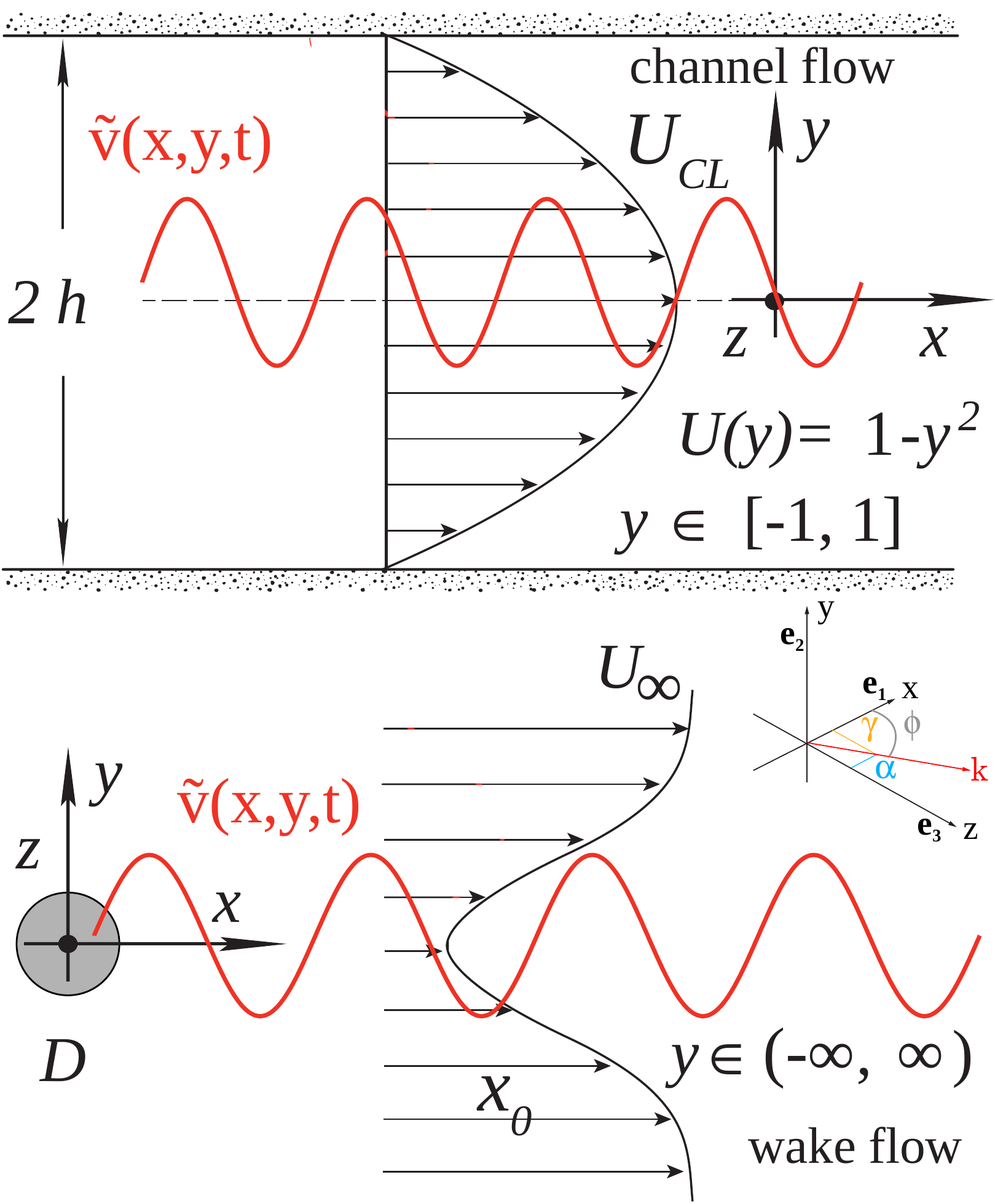}
\caption {Basic flows and perturbation scheme in the x-y physical plane. A Cartesian reference frame is adopted, with unit vectors $\mathbf{e_1}$, $\mathbf{e_2}$, $\mathbf{e_3}$ in the $x$, $y$, $z$ directions, respectively. \textcolor{black}{The base flow profiles are qualitatively represented in black with arrows.  A generic perturbation wave is represented in red. $\textbf{k}  = \alpha \mathbf{e_1} + \gamma \mathbf{e_3}$ is the wavenumber vector, $\phi$ is its angle with respect to the basic flow $\mathbf{U}=U(y)\mathbf{e_1}$. In the physical domain, the longitudinal space variable is unbounded for the channel flow and positive for the wake flow. However, due to the adoption of the near-parallel flow approximation, see section II, the domain for the plane wake is also seen as unbounded. The spanwise coordinate is unbounded for both flows. Note that the transversal computational domain width is defined through the parameter $y_f$. For the channel flow $y_f =\pm 1$, while for the wake flow $y_f$ is defined so that the numerical solution be insensitive to further extensions of the computational domain ($y_f = 20$ for short waves and $y_f$ up to $100$ for long waves). }
}
\label{Fig_1}
\end{figure}

\begin{align}
\dex \widetilde{u}+ \dey \widetilde{v} + \dez \widetilde{w} &= 0  \label{NS_lin_continuity}
\\
\dt \widetilde{u} + U\dex \widetilde{u}+\widetilde{v}U^{\prime} +  \dex \widetilde{p}&= \frac{1}{Re}
\nabla^2\widetilde{u}\hspace{20pt} \label{NS_mom1}
\\
 \dt \widetilde{v}+ U \dex \widetilde{v} +\dey \widetilde{p} &= \frac{1}{Re}
\nabla^2\widetilde{v}\hspace{20pt} \label{NS_mom2}
\\
 \dt \widetilde{w} + U \dex \widetilde{w} + \dez \widetilde{p}&= \frac{1}{Re} \nabla^2\widetilde{w}\hspace{20pt}
\label{NS_mom3}
\end{align}
\noindent where ($\widetilde{u}(x, y, z, t)$, $\widetilde{v}(x, y,
z, t)$, $\widetilde{w}(x, y, z, t)$) and $\widetilde{p}(x, y, z, t)$
are the components of the perturbation velocity and pressure,
respectively. The domain is $-\infty<x,z<\infty$, $-1<y<1$ for the channel flow, and $0<x<\infty$, $-\infty<z<\infty$, $-\infty<y<\infty$ for the wake. \textcolor{black}{No-slip boundary conditions for the channel flow imply vanishing perturbation at the walls, $\widetilde{u}=\widetilde{v}=\widetilde{w}=0$ as $y=\pm 1$, while decaying disturbances are considered for the wake flow, which is unbounded, so that $\widetilde{u},\widetilde{v},\widetilde{w}\to0$ as $y\to \infty$. Periodicity boundary conditions are used in both $x$ and $z$ directions. 
}
The channel  half-width $h$, and the body diameter $D$ are considered as reference external length scales. The reference velocity for the channel flow is the centreline velocity $U_{\rm CL}$, while in the wake case the free-stream velocity $U_f$ is considered. The reference time is the convective one.  Consequently, the flow control parameter is the Reynolds number, defined as $Re = U_{\rm CL} h/\nu$, for the channel flow, and as $Re = U_f D/\nu$ for the wake flow, where $\nu$ is the kinematic viscosity. The channel base flow is represented by the plane Poiseuille solution 
\begin{equation}
U(y)=1-y^{2}.
\label{U_channel}
\end{equation}
As a wake basic flow, we use the first two order terms of the Navier--Stokes  solution described in \cite{bt2002, tb03} and reported below:
\begin{equation}
U(y; x_{0}, Re)=C_{0}-\frac{1}{\sqrt{x_0}}C_{1}e^{-Re\ y^{2}/(4x_0)}\hspace{30pt}
\label{U_wake}
\end{equation}
\noindent where $C_{0}=1$ and $C_1=1.22 + 0.000067 Re^2$. \textcolor{black}{
The two-dimensional wake is a \textcolor{black}{spatially evolving} free flow. Leaving aside the near field, that is highly non parallel since it hosts the two symmetric countercirculating vortices that constitute the separation
region, the intermediate and long term wake is a near-parallel flow. The wake slowly becomes thicker according to a law which, at first order, scales as  $(x/Re)^{0.5}$. As representation of this steady subcritical flow we consider the previously mentioned  asymptotic expansion solution in inverse powers of $x$ which was also used in convective instability wake intermediate asymptotics analysys \citep{bt2006, tsb2006, stc10}. In particular, we consider the intermediate - far field well represented by  sections located in the interval  $x \in [5, \infty]$, in the range of Reynolds numbers $[20, 100]$. The basic flow is approximated by freezing it at three longitudinal stations placed at $x_0 = 10, 20, 50$. In so doing, the basic flow is parameterized with the downstream station $x_0$ and the Reynolds number. It is thus homogeneous in $x$ and $z$.
} The momentum equations Eq. (\ref{NS_mom1}-\ref{NS_mom3}) can be expressed in terms of the velocity-vorticity formulation:
\begin{flalign}\label{veloc_vort1}
\big[\big(\dt+U\dex \big)\nabla^2-U^{\prime \prime}\dex-\frac{1}{Re}
\nabla^4\big]\widetilde v &=0\\ \label{veloc_vort2}
\big[\dt +U\dex-\frac{1}{Re}\nabla^2\big]\widetilde{\omega_y} &=-U^{ \prime}\dez \widetilde{v}\\ 
\widetilde {\omega_y}&=\dez \widetilde u-\dex\widetilde w \label{veloc_vort3}
\end{flalign}
Eqs. (\ref{veloc_vort1},\ref{veloc_vort2},\ref{veloc_vort3}) are solved by means of a combined Fourier--Fourier (channel) and Laplace--Fourier (wake) transform in the plane normal to the basic flow profile. The transformation of a  generic variable $\widetilde g(x,y,z,t)$ reads:
\begin{equation}\label{transformation}
\hat{g}(y, t; \alpha, \gamma) = \int\limits_{-\infty} ^{+\infty} \int\limits_{\ \ 0, -\infty}
^{+\infty} \widetilde{g}(x, y, z, t) e^{-i \alpha x -i \gamma z} dx
dz 
\end{equation}
where $\alpha$ and $\gamma$ are the streamwise and spanwise wavenumbers, respectively.
For the sake of simplicity, in the present work only real wavenumbers are considered, even if for the wake flow the formulation allows to take into account spatially evolving waves. The wavenumber modulus is $k=\sqrt{\alpha^2+\gamma^2}$ and the wave obliquity angle is $\phi=tan^{-1}(\gamma/\alpha)$, see Fig. \ref{Fig_1}. The governing equations in the wavenumber space are thus formulated:

\begin{flalign} 
\big[\big(\dt+i\alpha U\big)\big(\ddey-k^2\big)- i\alpha U^{\prime \prime}\nonumber\\ -\frac{1}{Re}\big(\ddey-k^2 \big)^2 \big] \hat v &=0 \label{OS_IVP}\\
\big[\big(\dt+i\alpha U\big)-\frac{1}{Re}\big( \ddey-k^2 \big)\big]\hat\omega_y &=-i\gamma U^{\textcolor{black}{\prime}} \hat v
\label{Squire_IVP}
\end{flalign}
The streamwise and the spanwise velocity components, $\hat u=\frac{i}{k^2}(\alpha\frac{\partial \hat v}{\partial y}-\beta\hat{\omega_y})$, $\hat w=\frac{i}{k^2}(\beta\frac{\partial \hat v}{\partial y}+\alpha \hat{\omega_y})$ can be recovered from the continuity equation and from the vorticity definition.  The  boundary conditions associated to the system  \ref{OS_IVP}-\ref{Squire_IVP} for the channel flow are

\begin{equation}\label{BC_channel}
\hat v(\pm 1,t)=\dey \hat v(\pm 1,t)=\hat \omega_y(\pm 1,t)=0,
\end{equation}
while in the wake case we consider finite-energy harmonic velocity as $|y|\to\infty$, and vanishing vorticity in the free-stream, see also \cite{gs1978} and \cite{stc09}:
\begin{equation}\label{BC_wake}
\ddey \hat v=k^2\hat v,\hspace{10pt} \hat\omega_y=0\hspace{20pt}y\to\pm\infty, \forall t
\end{equation}
About the initial conditions, please, refer to section \ref{Sec4}.
The eigenvalues $\sigma=\sigma_r+i\sigma_i$ and the eigenfunctions of the eigenproblem associated to the system Eq. (\ref{OS_IVP}-\ref{Squire_IVP}) are computed by means of three different methods: a $5^{th}$-order Galerkin method based on Chandrasekhar expansions (see \cite{f2013}, and Appendix A) ,  a finite difference $4^{th}$-order scheme, \cite{dsct2015}, and a hybrid spectral collocation method based on Chebyshev polynomials \cite{sh2001}. A comparison of this techniques and the details of the first method are given in Appendix A and B. 

Two different numerical methods are used to solve the initial value problem: the method of lines, based on finite difference spatial discretization and a Runge-Kutta (2,3) temporal integration scheme, and the $5^{th}$-order Galerkin method  based on the  Chandrasekhar functions expansion cited above. The results obtained by the two approaches present a really good agreement. The computational domain is $[-y_f,y_f]$. For the channel flow $y_f=1$, while for the wake flow $y_f$ is defined so that the numerical solution be insensitive to further extensions of the computational domain size ($y_f = 20$ for short waves and $y_f$ up to $100$ for long waves). The solution computed by the initial-value formulation at long enough time shows excellent agreement with the behavior predicted by the modal analysis.  \\

In order to measure the growth of the perturbations, we define the kinetic energy density as 
\begin{equation}
e(t; \alpha, \gamma) = \frac{1}{4 y_f} \int_{-y_f}^{+y_f}
(|\hat{u}|^2
+ |\hat{v}|^2 + |\hat{w}|^2) dy.
\end{equation}
We then introduce the amplification factor, $G$, as the kinetic energy density normalized with respect to its initial value,
\begin{equation}
G(t; \alpha, \gamma) = e(t; \alpha, \gamma)/e(t=0; \alpha, \gamma).
\label{G}
\end{equation}
\noindent  Since the \textcolor{black}{temporal asymptotic} behavior of the linear perturbations is exponential, the temporal growth rate $r$ is defined \cite{cjj2003} as
\begin{equation}
r(t; \alpha, \gamma) = log(G)/(2t).
\label{tgr}
\end{equation}

The temporal evolution of the kinetic energy is given by the equation \citep{rh1993,sh2001}:
\begin{eqnarray}
\frac{de}{dt}=& &\frac{1}{k^{2}}\Im\int U'\bigg( \alpha\bar v \dey {\hat v}-\gamma\bar v \hat{\omega}_{y}\bigg)\ dy \nonumber \\ &-&\frac{1}{Re\ k^{2}}\int \bigg(|\ddey \hat v|^{2}+2k^2|\dey\hat v|^{2}+k^4|\hat v|^2\nonumber \\ &+& |\dey\hat {\omega}_{y}|+k^2| \hat{\omega}_{y}|^2\bigg)dy,
\label{qqqq}
\end{eqnarray}
where the bar indicates the complex-conjugate and $\Im$ is the imaginary part. While the dissipative term is always negative, the convective term has undetermined sign and can be responsible for kinetic energy growth.

\noindent The frequency of the perturbation $\omega$ is
defined as the temporal derivative of the unwrapped wave phase $\theta(y, t; \alpha, \gamma)$, at a specific spatial point along the $y$ direction. The wrapped phase,

\begin{equation}\label{th}
\theta_w(y, t; \alpha, \gamma) = arg(\hat{v}(y, t; \alpha, \gamma)),
\end{equation}
\noindent is a discontinuous function of $t$ defined in $[-\pi,+\pi]$, while the unwrapped phase $\theta$ is a continuous function obtained by introducing a sequence of $2 \pi$ shifts on the phase values in correspondence to the periodical discontinuities. In the case of the wake we use as reference transversal observation point $y_0 = 1$ or $y_0=5$, and in the case of the channel flow the point $y_0 = 0.5$. The frequency \cite{stc09} is thus

\begin{equation}\label{fr}
\omega(t; y_0, \alpha, \gamma) = |d \theta(t; y_0, \alpha, \gamma)|/dt.
\end{equation}

\noindent The phase velocity is defined as

\begin{equation}\label{c}
{\bf c}=(\omega/ k) \hat{\textbf{k}},
\end{equation}

\noindent where $\hat{\textbf{k}} = \cos(\phi){\bf e_1}+\sin(\phi){\bf e_3}$ is the unit vector in the direction of \textbf{k}.

\section{Dispersive to non-dispersive behavior in the long term}
\begin{figure*}
\begin{center}
\includegraphics[width=.9\textwidth]{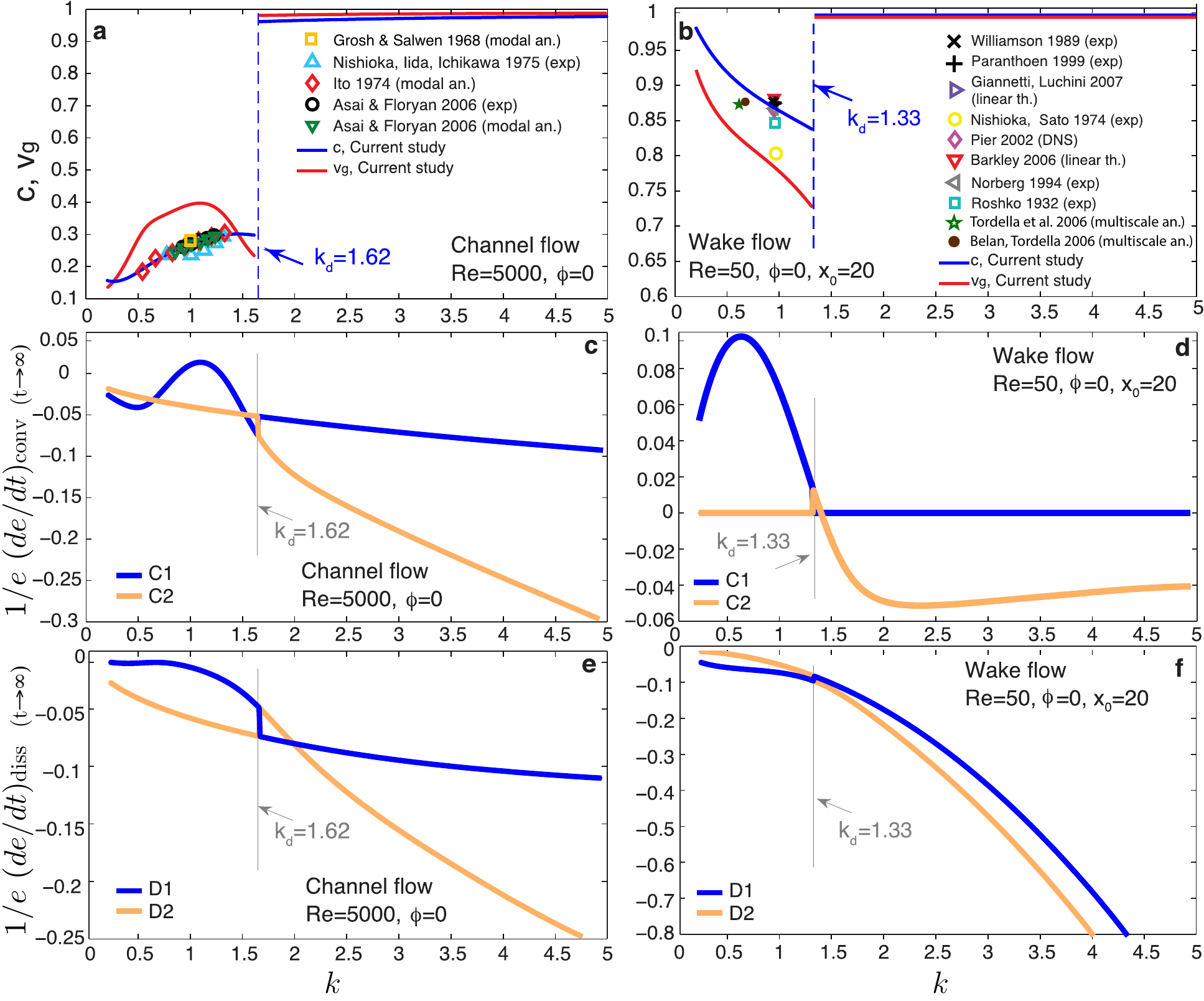}
\end{center}\vspace{-10pt}
\caption {\textcolor{black}{Dispersion relations for longitudinal waves ($\phi=0$) in both the channel and wake flow. Results from eigenvalue analysis, in particular from least stable mode characteristics. (a,b) Phase and group velocity as a function of the wavenumber and comparison with other studies. The phase velocity is represented with the blue line, while the group velocity line is red. For more info on the relation $\omega(k)$ see also Figs. S1 and S2 in the SM. The wavenumber range $[0.2,5]$ is discretized with $\Delta k=0.005$. $k_d$ indicates the threshold value that separates the dispersive waves from  non-dispersive waves. 
(a) Channel flow with $Re=5000$, $k_d=1.62$ and comparison with other studies, see Grosh \& Salwen 1968 \cite{gs1968}, Ito 1974 \cite{ito1974}, Nihsioka et al. 1975 \cite{ni1975}, Asai \& Florian 2006 \cite{asai2006}. (b) Wake flow, $Re=50$ and $x_0=20$. Here $k_d=1.32$. A comparison is given with the phase velocity values found by Roshko 1974 \cite{r1954}, Nihsioka \& Sato 1974 \cite{ns1974}, Norberg 1994 \cite{n1994}, Williamson 1989 \cite{w89}, Parantho\"en et al. 1999 \cite{p1999}, Pier 2002 \cite{p2002}, Barkley 2006 \cite{b2006}, Belan \& Tordella 2006 \cite{bt2006}, Tordella et al. 2006 \cite{tsb2006}, Giannetti \& Luchini 2007 \cite{gl2007}. Wherever not available, the wavelength has been set equal to the mean of the values measured by Parantho\"en et al. ($\lambda = 7 D$) and Williamson ($\lambda = 7.15 D$). (c-f)  Convective and dissipative energy variation terms as defined in eq. \ref{dedt} as a function of the wavenumber, for the two least-stable eigenmodes of the left and right branch of the spectrum. Blue lines represent the least-stable solution. For the channel flow at Re = 5000, the evolution of two dispersive, one of which centered on $k_d$, and one \textcolor{black}{non-dispersive} wave packets is shown in three movies in the SM online \cite{supplementary}.}}
\label{dispersion_relations}
\end{figure*}

The focus of this section is on the \textcolor{black}{long-term temporal behavior} of small amplitude traveling waves, and on the relation between the frequency of the least stable eigenvalue and the wavenumber. The wavenumber is within a wide range $0.2<k<20$, with a small step $\delta k = 0.005$. This allows to highlight a transition from  dispersive to \textcolor{black}{non-dispersive} behavior. This transition is also useful to understand the temporal evolution of general arbitrary initial perturbations, which is the object of the following sections. The results here presented, see Tables I and II and Fig. \ref{dispersion_relations}, have been compared with literature data obtained by means of different methods of investigation, in particular laboratory, modal and Initial Value Problem analysis. Least stable eigenvalue frequencies are deduced from the computation of the eigenvalue spectra at constant wavenumber, waveangle and Reynolds number.

\noindent Figure \ref{dispersion_relations} (a,b) shows the dependence of the phase velocity $c$ on the polar wavenumber $k$ in the range $k \in [0.2, 5]$. In both the flows we observe the existence of a threshold wavenumber that we indicate with $k_d$ where a step variation of the phase velocity occurs. For $k > k_{d}$ the phase velocity is approximately constant and is nearly equal to the group velocity, a fact which highlights a non-dispersive behavior. For $k<k_{d}$ instead, $c$ depends on  $k$ in a different way  from the group velocity and the behavior is \textcolor{black}{dispersive}. 

\textcolor{black}{
In other word, inside the range of wavelengths that can be hosted in the system, there is a wavenumber threshold above which the least stable mode is always on the tip of the right branch of the eigenvalue spectrum (about the branched structure of the eigenvalue spectrum, see for example \cite{cjj2003, sh2001} and panels d and h of figure 4 below). By increasing the wavenumber, the frequency of the least stable mode increases in a proportional way, which yields \textcolor{black}{non-dispersiveness}. Viceversa, below this threshold, the least stable mode is located on the tip of the left branch. By varying the wavenumber one sees that this least stable mode frequency does not vary in a way that is proportional to the wavenumber variation, which yields dispersivity. In terms of phase and group velocity this threshold yields a sharp transition because the tips of the left and right branches of the eigenvalue spectrum have substantially different frequencies, see again Fig. \ref{fig_4_new}, panels d,h.
It should be considered that during the transient part of the evolution of the sum of a number of perturbations with different wave numbers (for example a wave packet) one can better perceive the contribution of the least stable modes for each wavenumber, which in turn can be associated to different spectral branches and have different frequencies and growth or decay factors. In this last case, in the long term, the solution will mainly  contain the less stable components for each specific wave number inside. These least stable modes will disperse if their wavelength is below the $k_d$ threshold slightly sensitive to the Reynolds number, see tables I and II, and that the opposite is true for wavelengths above the threshold. }

\textcolor{black}{A question naturally arises: why does the set of least-damped waves become \textcolor{black}{non-dispersive} above a given $k$? In this regard, we like to highlight the following.
	As already said, it can be observed that for $k < k_d$, the least-stable Orr-Sommerfeld eigenvalue is located on the tip of the left branch of the eigenvalues spectrum. \textcolor{black}{As described in further detail} in Section V (see figures 6 and 7), the related velocity  eigenfunction is a \textit{shear mode}: the highest perturbation variations ($d\hat v/dy$) are located where the base flow vorticity $dU/dy$ is high (in the shear region). This region is located close to the walls for the plane Poiseuille flow, while it is close to the inflection points in the wake flow. We refer to these perturbations as \textit{wall modes} for channel flow, and \textit{in-wake modes} for the wake. 	Vicevera, for $k > k_d$, the least-stable eigenvalue is located on the tip of the right branch and the dominant perturbation solution is an \textit{external mode}: in this case the perturbation varies rapidly where base flow has low vorticity (out of the high-shear region). In synthesis, when the wavelength of a perturbation is short enough ($k>k_d$) the dominant mode variation $d\hat v/dy$ is confined in a region of low base flow vorticity. In this case, note that the dispersion relation for the least-damped modes is close to the dispersion relation for a uniform base flow which is always \textcolor{black}{non-dispersive} (see \cite{sh2001}, chap 3.2). Here,  we refer to \textit{central modes} for the channel flow, and \textit{out-of-wake modes} for the wake flow.}

\textcolor{black}{Coming back to the comparison with data in literature, one can see that the agreement between laboratory data as well as numerical modal analysis and our work is very good. In particular, as regards the channel flow, we observe a good agreement with the works of Grosh \& Salwen 1968 \cite{gs1968}, Ito 1974 \cite{ito1974}, Nishioka et al. 1975 \cite{ni1975},  and Asai \& Floryan 2006 \cite{asai2006} (see panel (a) of Fig. \ref{dispersion_relations}). Grosh \& Salwen give the least stable eigenvalues in the range of $\alpha$ [1 - 2.4] , $\phi=0$, and in the range [5 - 25000] of the Reynolds number (under our normalization). 
Asai and Floryan investigated experimentally the  effects of wall corrugation on the stability of wall-bounded shear flows and found a good agreement with the Orr-Sommerfeld theory coupled to roughness in the form of a single Fourier harmonic and in the form of spanwise grooves with rectangular and triangular shapes \cite{floryan2005}.
}

\textcolor{black}{For the wake flow}, it should be observed that in literature data are mostly focused on the value of the shedding frequency. Our references include the works by Roshko 1954 \cite{r1954}, Nishioka \& Sato 1974 \cite{ns1974}, Williamson 1989 \cite{w89}, Norberg 1994 \cite{n1994}, Paranth\"oen et al. 1999 \cite{p1999},  Pier 2002 \cite{p2002}, Barkley 2006 \cite{b2006}, Belan \& Tordella 2006 \cite{bt2006}, Tordella et al. 2006 \cite{tsb2006}, and Giannetti \& Luchini 2007 \cite{gl2007}. The agreement is good also in this case. Indeed the relative error with respect to the measurements by Williamson and Paranth\"oen et al. is about $3.3\%$. \textcolor{black}{We would like to remind the reader} that wake data in literature usually give information on the frequency at which vortex shedding takes place but don't give information about the wavenumber of the shedding. Thus, to address this lack of information, we decided to associate to the frequency values reported in literature the average of the wavenumber values measurements currently available for $Re = 50$, that is Williamson \cite{w89} and Paranth\"oen \cite{p1999}. So that, leaving aside the two points related to results from near-parallel multiple scale convective analysis \cite{bt2006, tsb2006}, where wavelength and frequency are determined along the streamwise direction, all the points in figure 2 b have been  superposed on the same wavenumber.
%


\textcolor{black}{Coming back to Table I and II, on can see that the threshold $k_d$ is a function of the Reynolds number and the wave angle but for the wake flow - which is a weakly streamwise evolving flow - it is also a function of the streamwise station $x_0$. In particular, for the channel flow, values of $k_d$ have been computed in the range  $Re\in[1000,8000]$ and $\phi\in[0,\pi/3]$. 
We observe that $k_d$ decreases as $Re$ and increases as the wave angle. The $k_d-Re$ trend is opposite in the wake case, see Table \ref{kd_wake} 
where for  $Re\in[20,100]$  the dependence on the flow station $x_0$ is shown ($x_0=[10,20,50]$).} 
\textcolor{black}{It is observed that in the case of the channel flow, the system tends to become non-dispersive at all wavenumbers as Re tends to infinite. As regards to the wake,  this trend is reversed. This may not be intuitive but one can consider that with increasing Re, even though the wake narrows, the shear intensity increases and the range of  dispersive waves widens. This supports the fact that the base vorticity intensity plays a key role in the dispersion. }

\begin{table}[]
\caption{\label{kd_channel}
Values of the dispersive regime threshold wavenumber $k_d$ for the channel flow, for different Reynolds number and  obliquity angles. The uncertainty on $k_d$ due to the discretization is $\pm 0.005$.  }
\begin{ruledtabular}
\begin{tabular}{ccccc}
\multicolumn{5}{c}{Channel flow}\\
\textbf{Re}&  $\ \mathbf{\bm\phi=0}\ $ & $\mathbf{\bm\phi=\bm\pi/6}$ & $\mathbf{\bm\phi=\bm\pi/4}$ &$\mathbf{\bm\phi=\bm\pi/3}$  \\ 
\hline
\textbf{1000} &    2.071  &  2.111 &    2.168  & 2.256  \\ 
\textbf{2000} &    1.883  &  1.922 &    1.979  & 2.073  \\ 
\textbf{3000} &    1.764  &  1.803 &    1.866  & 1.960  \\ 
\textbf{4000} &    1.686  &  1.725 &    1.784  & 1.878  \\ 
\textbf{5000} &    1.623  &  1.662 &    1.721  & 1.815  \\   
\textbf{6000} &    1.576  &  1.615 &    1.670  & 1.765  \\  
\textbf{7000} &    1.536  &  1.568 &    1.627  & 1.720  \\  
\textbf{8000} &    1.497  &  1.536 &    1.589  & 1.682  \\  
\end{tabular}
\end{ruledtabular}
\end{table}

\begin{table}[]
\caption{\label{kd_wake}
Values of the dispersive regime threshold wavenumber $k_d$ (uncertainty of $\pm0.005$) for the wake flow, for different Reynolds number,  obliquity angles and streamwise station $x_0$. }
\begin{ruledtabular}
\begin{tabular}{ccccc}
\multicolumn{5}{c}{\rule{0 pt}{10pt}$x_0=10$}\\   
\textbf{Re} &  $\ \mathbf{\bm \phi=0}\ $ & $\mathbf{\bm\phi=\bm\pi/6}$ & $\mathbf{\bm\phi=\bm\pi/4}$ &$\mathbf{\bm\phi=\bm\pi/3}$  \\
\hline
\textbf{20}  & 1.073 & 1.032 & 0.977 & 0.868           \\
\textbf{30}  & 1.373 & 1.333 & 1.276 & 1.154           \\
\textbf{40}  & 1.636 & 1.598 & 1.534 & 1.407           \\
\textbf{50}  & 1.875 & 1.835 & 1.772 & 1.640           \\
\textbf{60}  & 2.091 & 2.052 & 1.989 & 1.862           \\
\textbf{70}  & 2.293 & 2.253 & 2.195 & 2.068           \\
\textbf{80}  & 2.485 & 2.448 & 2.390 & 2.264           \\
\textbf{90}  & 2.663 & 2.628 & 2.575 & 2.453           \\
\textbf{100} & 2.837 & 2.802 & 2.749 & 2.633           \\
\hline
\multicolumn{5}{c}{\rule{0 pt}{10pt}$x_0=20$}\\   
\textbf{Re}&  $\ \mathbf{\bm\phi=0}\ $ & $\mathbf{\bm\phi=\bm\pi/6}$ & $\mathbf{\bm\phi=\bm\pi/4}$ &$\mathbf{\bm\phi=\bm\pi/3}$  \\
\hline
\textbf{20}  & 0.758 & 0.735 & 0.696 & 0.617           \\
\textbf{30}  & 0.974 & 0.946 & 0.903 & 0.817           \\
\textbf{40}  & 1.158 & 1.131 & 1.087 & 0.997           \\
\textbf{50}  & 1.326 & 1.298 & 1.251 & 1.161           \\
\textbf{60}  & 1.478 & 1.451 & 1.408 & 1.318           \\
\textbf{70}  & 1.623 & 1.596 & 1.553 & 1.463           \\
\textbf{80}  & 1.756 & 1.733 & 1.689 & 1.604           \\
\textbf{90}  & 1.885 & 1.858 & 1.819 & 1.737           \\
\textbf{100} & 2.001 & 1.983 & 1.944 & 1.862           \\
\hline
\multicolumn{5}{c}{\rule{0 pt}{10pt}$x_0=50$}\\   
\textbf{Re}  &  $\mathbf{\bm\phi=0}$ & $\mathbf{\bm\phi=\bm\pi/6}$ & $\mathbf{\bm\phi=\bm\pi/4}$ &$\mathbf{\bm\phi=\bm\pi/3}$  \\
\hline
\textbf{20}  & 0.482 & 0.461 & 0.441 & 0.401           \\
\textbf{30}  & 0.615 & 0.593 & 0.573 & 0.522           \\
\textbf{40}  & 0.732 & 0.714 & 0.684 & 0.633           \\
\textbf{50}  & 0.840 & 0.815 & 0.795 & 0.734           \\
\textbf{60}  & 0.935 & 0.916 & 0.886 & 0.835           \\
\textbf{70}  & 1.026 & 1.007 & 0.977 & 0.926           \\
\textbf{80}  & 1.112 & 1.088 & 1.068 & 1.007           \\
\textbf{90}  & 1.189 & 1.169 & 1.148 & 1.098           \\
\textbf{100} & 1.267 & 1.249 & 1.229 & 1.179           \\
\end{tabular}
\end{ruledtabular}
\end{table} 

Finally, by considering the kinetic energy equation for two-dimensional perturbations,
\begin{gather}
\frac{de}{dt}=\underbrace{\frac{1}{k^{2}}\Im\int U'\bigg( \alpha\bar v \dey {\hat v}\bigg)\ dy}_{Convection} \nonumber \\ \underbrace{-\frac{1}{Re\ k^{2}}\int \bigg(|\ddey \hat v|^{2}+2k^2|\dey\hat v|^{2}+k^4|\hat v|^2 \bigg)dy}_{Dissipation},
\label{dedt}
\end{gather}
we compared the convective and the dissipative terms for the least stable mode on the left and right branches of the spectrum respectively, see Fig. \ref{dispersion_relations} panels (c,d,e,f). For simplicity, we have considered only symmetric modes but it can be shown that the results do not change when  antisymmetric modes are considered. In  Fig. \ref{dispersion_relations}, $C1$ and $D1$ indicate, for the least stable mode, the contribution of the convective and viscous terms of equation (\ref{dedt}) to the normalized kinetic energy temporal rate of change, respectively. $C2$ and $D2$ indicate the contribution given by the second-last least damped mode, which for $k>k_d$ belongs to the P-family of eigenvalues for the Poiseuille flow, and to the continuous branch for the wake flow. It can be observed that by means of an interchange of modes taking place at $k=k_d$, the system, beyond this threshold, settles to the mode that minimizes the normalized kinetic energy loss.  
In synthesis, beyond  $k_d$ the traveling waves are stable and \textcolor{black}{non-dispersive}, and the rate of kinetic energy loss associated to the convective effects is independent (wake case) or very mildly dependent (channel case) on the wavenumber.  Cases of mode exchange for shear or stratified flows with complex dispersion relationship have been reported and analyzed by other authors (see, for instance, \cite{sg1981,scg1980} for the Poiseuille pipe flow and and \cite{gh1980} for the stratified Couette flow). For a review, see \cite{craik1985}, sections 2.2 and 2.6. It should be noted that the mode exchange observed here for the plane channel flow went mostly unnoticed up to now. In fact, we were able to only find a short reference inside Fig 3.8b in \cite{sh2001}, page 71. Possibly, because it takes place in the very stable region above the neutral curve of the $k$-Re plane, that is, a region not of great interest when the focus is on instability. 

In the next sections the transient evolution of small perturbations is considered. We show that the threshold $k_d$ has a key role in their temporal evolution.	

 
\section{Phase velocity transient dynamics}\label{Sec4}

Many studies on shear flows have shown the importance of the early time dynamics, that in principle can lead to large transient energy growth long before the exponential mode becomes dominant \citep{bf1992, cd90, hlj1993, hr1994, l1996, ljjc1999, bb2009}. Transient dynamics offers a  variety of different behaviors and phenomena which are not easy to predict \emph{a priori}. 
In the literature transient dynamics is generally analyzed in terms of energy amplification factor, while lesser attention has been paid to the phase speed evolution. \textcolor{black}{
However, the temporal evolution of the phase speed can lead to interesting considerations and useful information. In particular, we can understand why and when different time scales appear inside the transient (see figures 4-7 below) and how these scales are linked to the features of the eigenvalue spectrum (for instance the spectral width). In fact, in the following, it can be observed that the sequence of different transient phases leading to the final temporal asymptote can be better observed through the phase speed rather than through the kinetic energy growth factor.}

In the context of our formulation, proper initial conditions on $\hat v$ and $\hat \omega_y$ have to be associated to  the system of equations \ref{OS_IVP}-\ref{Squire_IVP}. We have computed the phase velocity through the transversal velocity component $\hat v$, see Eqs. \ref{th}-\ref{c} but we could have considered any component of velocity and vorticity. 
Moreover Eq. \ref{OS_IVP}  is homogeneous and it can be demonstrated that the eventual introduction of an initial transversal vorticity does not actually affect the perturbation temporal evolution, see Ref. \cite{stc09}. For this reason, the initial vorticity $\hat \omega_{y}(0,y)$ is set to zero for all the simulations here considered.
\textcolor{black}{As for the velocity field, we consider four different initial conditions,  $ \hat{v} (0, y)$ that we selected on the basis of the symmetry and the location of the perturbation with respect to the higher or lower vorticity region in the basic flow}.

For a fixed wavenumber and Reynolds number, the combination of these two attributes produces a different level of excitation of the least stable subset of eigenvalues.
In particular, for the  base flows here considered, the excited least stable eigenvalues can belong to both the left and the right branches of the spectrum, see panels (d,h) in Fig. \ref{fig_4_new}.
After a very first stage where the contribution from the most damped eigenmodes disappears very quickly, and since the growth/decay rates in the subset of the least stable eigenvalues can be vey close, a second stage may set in where a sort of \textcolor{black}{intermediate temporal asymptotics} lasting hundreds or more time scales intervenes. This intermediate term \textcolor{black}{leads to the final state}, where the phase velocity takes the value corresponding to the last symmetric (or antisymmetric) eigenvalue, depending  on the parity of the initial condition, see  Fig.  \ref{fig_4_new}(d,h).

\begin{figure}
\begin{center}	
\includegraphics[width=0.95\columnwidth]{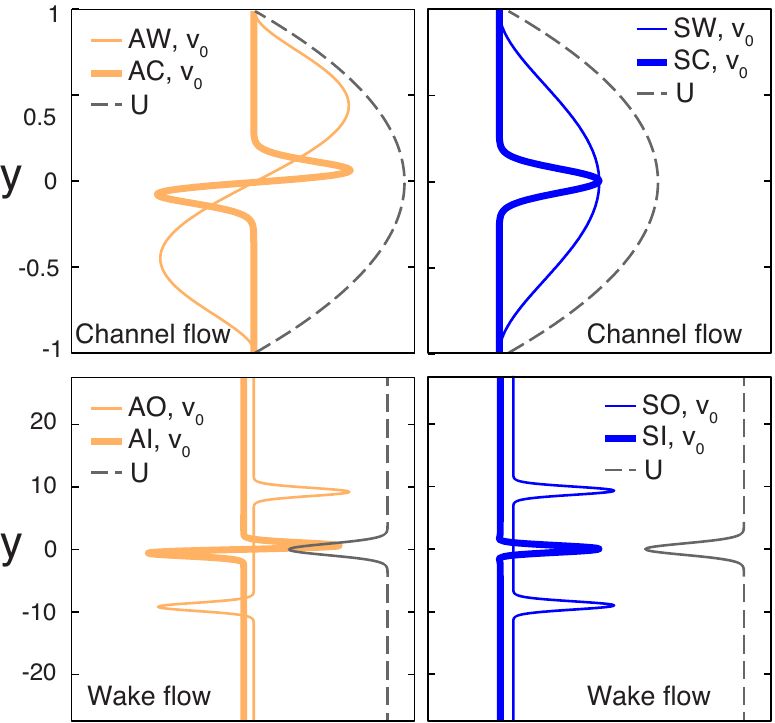}
\end{center}
\caption {Initial velocity disturbances $\hat v_{0}$ and base flows (dashed lines). The amplitudes have been scaled for clarity. Top panels: channel flow. Bottom panels: wake flow. The blue lines represent symmetric disturbances (right panels), while the antisymmetric are in yellow (left panels). Thick lines indicate central initial disturbances, thin curves represent the lateral ones.}
\label{fig_3_new}
\end{figure}

\begin{figure*}
\includegraphics[width=0.9\textwidth]{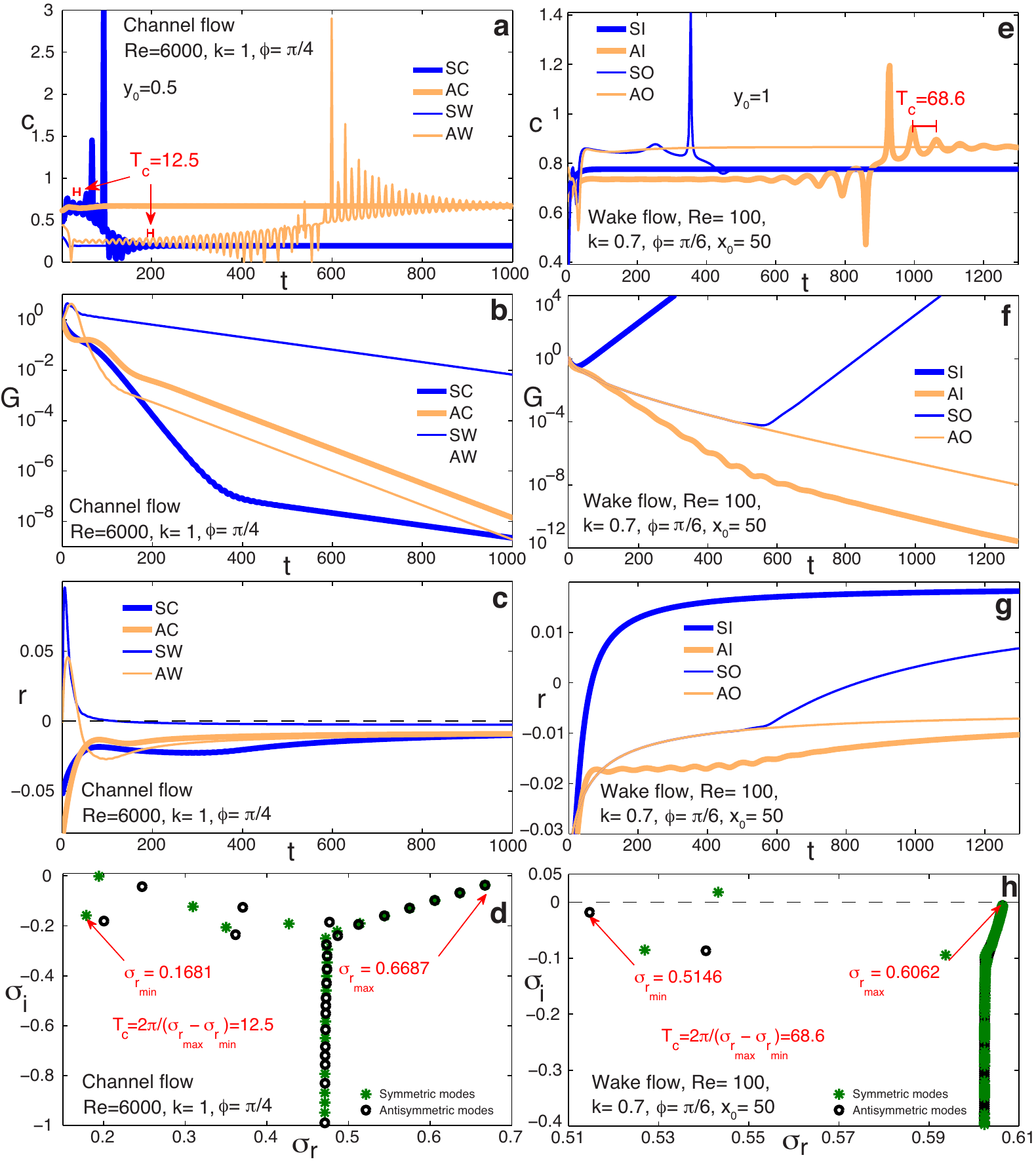}
\caption{Temporal evolution of perturbations in terms of the phase velocity $c$ (a,e), the amplification factor $G$ (b,f), and the temporal growth rate $r$ (panels c,g), for different initial conditions. The eigenvalues spectra are shown in panels (d, h). Left column: channel flow with $Re=6000$, $\phi=\pi/4$ and $k=1$. Right column: wake flow with $Re=100$, $\phi=\pi/6$ and $k=0.7$. The phase velocity is computed at $y_0=0.5$ for the channel and at $y_0=1$ for the wake flow (NB: the reference lengths are the channel half-width and the cylinder diameter). The quantity $T_c$, see (a) and (e), indicates the temporal periodicity of the phase velocity fluctuations observed in the early and intermediate term and corresponds to the ratio $T_c=2\pi/[\sigma_{r_{max}}-\sigma_{r_{min}}]$, as shown in panels (d) and (h).}
\label{fig_4_new}
\end{figure*}

The exact expressions of the initial conditions we use are given in Table \ref{ic_tab}  and they are shown for both base flows in Fig. \ref{fig_3_new}. In synthesis, we considered four different initial conditions for both the plane wake and channel flow. For the wake flow they will be named SI, AI, SO, AO (S=symmetric, A=antisymmetric, I=inside-wake, O=out-of-wake, see bottom panels of Fig. \ref{fig_3_new}). For the channel, which is bounded, we name them as SC, AC , SW, AW (C=central, W=wall, see top panels of Fig. \ref{fig_3_new}). We recall that the for the channel flow the high shear region is located near the wall while for the wake flow it is located in the central region of the open domain that computationally is typically from 20 to 100 times larger than the base flow.

\vspace{5pt}

\begin{table}[]
\caption{Initial conditions imposed on the velocity, $\hat v_0=\hat v(0,y)$. For the channel flow they are named SC/AC: symmetric/ antisymmetric and central. SW/AW: symmetric/antisymmetric and wall-located (the initial condition has large variations close to the channel walls). For the wake flow, SI/AI: symmetric/antisymmetric and inside-wake. SO/AO: symmetric/antisymmetric and out-of-wake (see Fig.  \ref{fig_3_new}).}
\label{ic_tab}
\begin{ruledtabular}
\begin{tabular}{ccc}
I.C. & Wake flow & Channel flow \\
\hline
SI/SC & $\hat v_0 =e^{-{y^2}}\cos(y)$ & $\hat v_0 =e^{-\frac{y^2}{0.01}}\cos(3y)$\\
AI/AC & $\hat v_0 =e^{-{y^2}}\sin(y)$ & $\hat v_0=e^{-\frac{y^2}{0.01}}\sin(3y)$\\
SO/SW & $\hat v_0=e^{-(y-10)^2}+e^{-(y+10)^2}$ & $\hat v_0=(1-y^2)^2$\\
AO/AW & $\hat v_0=e^{-(y-10)^2}-e^{-(y+10)^2}$ & $\hat v_0=y(1-y^2)^2$\\ 
\end{tabular}
\end{ruledtabular}
\end{table}

An overview of possible transients is presented in Fig. \ref{fig_4_new} where the evolution of three-dimensional perturbations is in terms of amplification factor $G$, phase velocity $c$, and temporal growth rate $r$.  The transient of perturbations with initial conditions as in Table \ref{ic_tab} are shown for both the base flows and for $k<k_d$. The results obtained by the IVP are compared with the eigenvalue spectra given by the modal analysis. The trends for small wave numbers are shown here, as they could be more easily observable in the laboratory. However, in the summary scheme of Fig.  \ref{schema}  the transient behavior of disturbances with wavenumbers larger than $k_d$ is shown. 
For the two kind of shear flows, the amplification factors and the temporal growth rates are reported in panels (b,c,f,g) of Fig. \ref{fig_4_new} . For the wake flow, we can observe that with an initial condition of the kind SI  -  symmetric and localized inside the base flow shear region - and  AO - antisymmetric external to the base flow shear region - the long term exponential behavior is reached after few temporal scales. On the contrary, with the other kind of initial conditions - SO and AI - the transients can last up to many hundreds of time units. In these cases quite far along within the transient the phase velocity abruptly changes its value, see panels (e). For the wake,  initial perturbations SO slow down, that is the phase velocity decreases after the jump. Perturbations AI, instead, experience an acceleration.
\\The same  behavior is observed in the case of the channel flow, but here the role of the initial conditions is reversed. In fact, SW and AC   show an exponential trend after few time scales, while SC and AW have a long transient characterized by the presence of phase velocity jumps (see Fig.  \ref{fig_4_new}a).  Also in this case, the phase velocity of the symmetric perturbation  jumps to a smaller value  while in the antisymmetric case it shifts to an higher value. The different behavior of the two base flows can be traced back to the structure of the eigenvalue spectra and the associated eigenmodes. Indeed, it is known that for the channel flow the eigensolutions of the left branch (A-branch) are wall-modes while those of the right branch (P-branch) are central modes. The opposite is true for the wake flow, whose spectra are made of a discrete set (left branch) of inside-wake modes, and a continuous branch of out-of-wake modes. As for the channel case, both branches may contain symmetrical and antisymmetric eigenfunctions, see panels (d,h) of Fig.   \ref{fig_4_new}.

As already reported in \cite{stc09}, besides the two temporal scales associated to the value of the phase velocity before and after the jumps, we observe a further periodicity, $T_c$, related to the temporal modulation of the phase velocity during the early and intermediate terms, see Fig. \ref{fig_4_new}(b,f). In this study we tried to relate this periodicity with the structure of the eigenvalue spectrum and verified that it is inversely proportional to the width of the range of frequencies given by the modal theory, $T_c=2\pi/[\sigma_{r_{max}}-\sigma_{r_{min}}]$.
\\ \indent The system presents  other two temporal scales: the external scale related to the base flow and the length of the transient (which can be determined by observing the time instant beyond which the growth rate $r$, and the angular frequency $\omega$, are both constant). Therefore, for a given wavenumber, from the phase speed transients it is possible to observe five different time scales.

\begin{figure}
\begin{center}
\includegraphics[width=0.9\columnwidth]{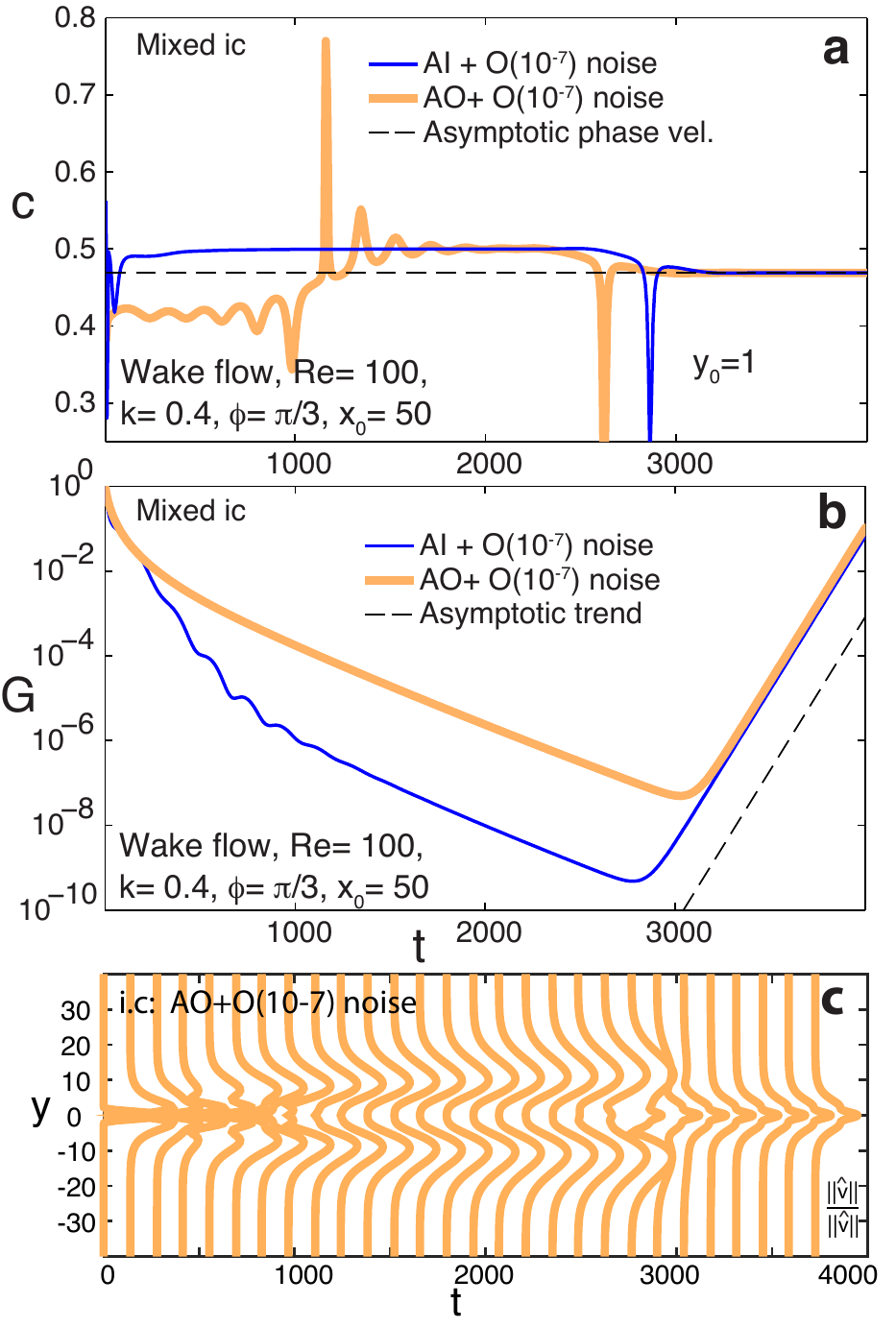}
\end{center}
\caption {Temporal evolution of mixed symmetric-antysimmetric initial conditions for the wake flow with $Re=100$, $k=0.4$, $\phi=\pi/3$ and $x_0=50$. A small uniform random noise of magnitude $O(10^{-7})$ is added to the initial conditions AI and AO in order to simulate the transient life of a general initial disturbance. (a) Growth rate. (b) Phase velocity. The dashed line represents the asymptotic trend given by the least stable Orr-Sommerfeld eigenvalue. More information about the solution is shown in Fig. S3 in the SM online \cite{supplementary}. (c)  Temporal evolution of the velocity profile modulus for the noisy AO initial condition. The profiles are normalized with respect to their peak value.  }
\label{noise}
\end{figure}

Fig.\ref{fig_4_new} shows that symmetric and antisymmetric perturbations do not have the same \textcolor{black}{temporal asymptotic} behavior. More precisely, symmetric perturbations are always less or equally stable than the antisymmetric ones. This means that if one considers mixed initial conditions, the symmetrical component will always prevail. Of course, in laboratory experiments, one cannot assume the precise symmetry because residual disturbance cannot be  completely suppressed. As a consequence, the phase velocity that can be eventually observed is that given by the symmetric part of the initial condition. In an attempt to reproduce what might happen in the laboratory, one can build up a mixed initial condition with prevailing antisymmetric component, by adding to the AI and the AO conditions a small noise which models the presence of a small symmetric part. The evolution of such a disturbance is shown in Fig.  \ref{noise}: after an early transient, the phase velocity experiences a jump (as it happens for AI and AO) and then the solution keeps the antisymmetric shape until the symmetric part becomes dominant. The \textcolor{black}{temporal asymptote} is eventually reached ($t<2900$) and it is announced by a second frequency jump. In this case the solution is a symmetric growing mode. 

\section{Near-similarity of velocity perturbations}
\begin{figure}
\begin{center}
\includegraphics[width=.8\columnwidth]{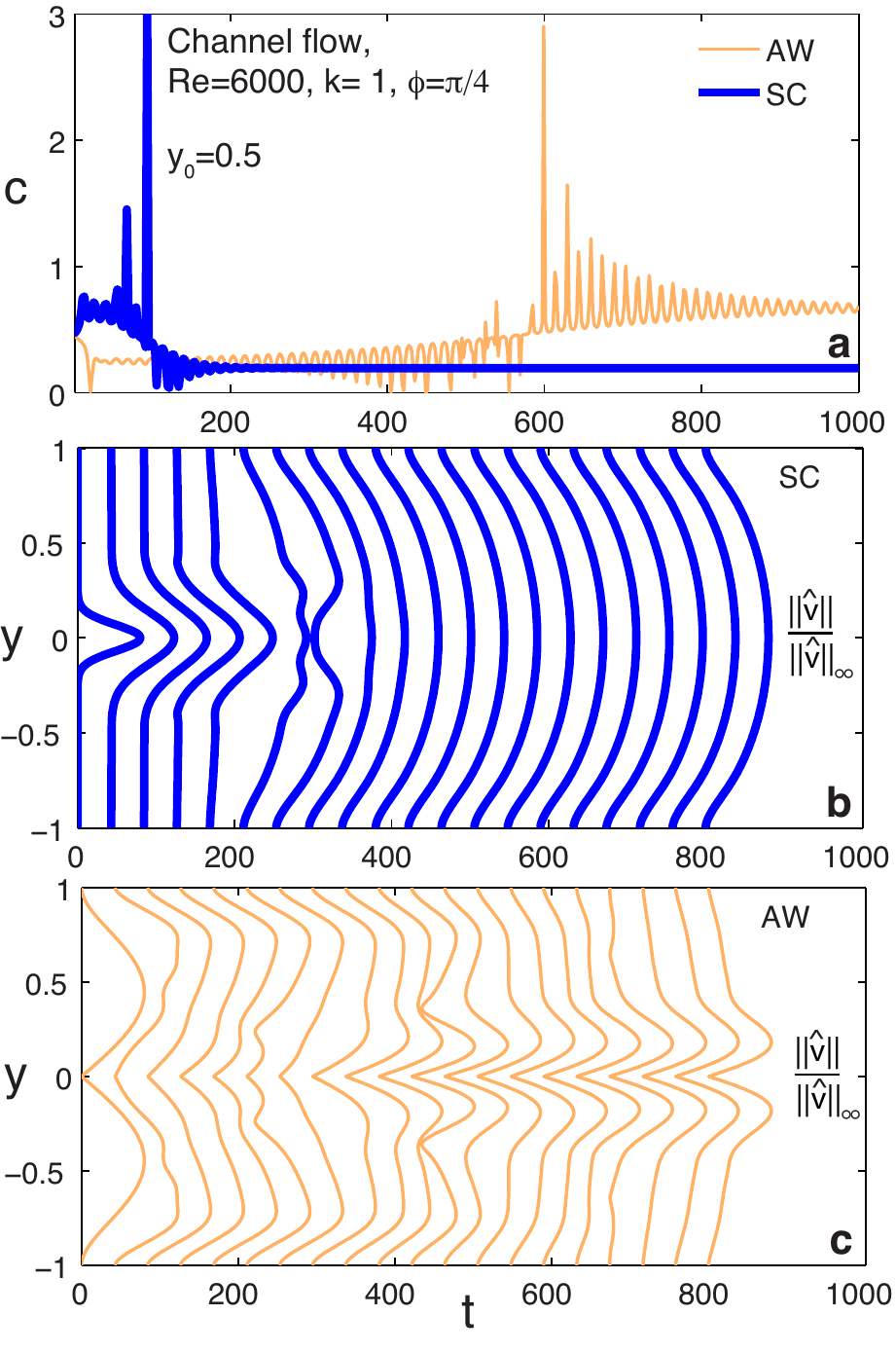}
\end{center}
\caption {Channel flow, $Re=6000$, $k=1$, $\phi=\pi/4$, initial conditions AW (yellow, thin curves) and SC (blue, thick curves). (a) Temporal evolution of the phase velocity, computed at $y_0=0.5$. A jump is observed at $t\approx100$ for SC and at $t\approx 600$ for AW.  (b) Evolution of the velocity profile modulus for the SC initial condition. The profiles are normalized with respect to their peak value. (c) Evolution of the normalized velocity profiles for the case of AW initial condition. The phase velocity shifts indicate a substantial change in the velocity profiles. This transition can be more or less abrupt depending on the simulation parameters, and  the regions before and after the jump can show a nearly self-similar behavior. }
\label{prof_c}
\end{figure}

\begin{figure}
\begin{center}
\includegraphics[width=.8\columnwidth]{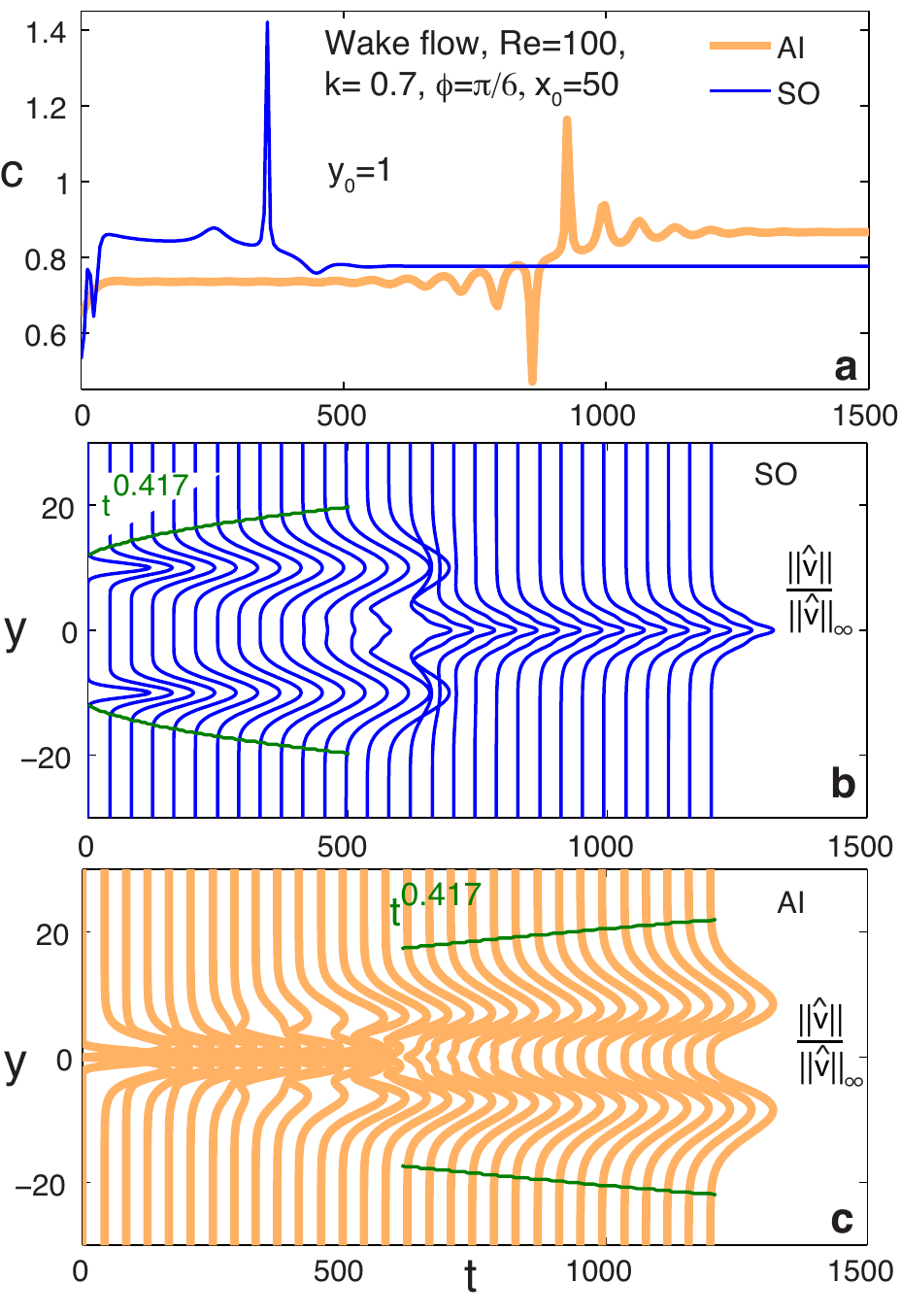}
\end{center}
\caption {Wake flow, $Re=100$, $k=0.7$, $\phi=\pi/6$, $x_0=50$, initial conditions AI (yellow, thick curves) and SO (blue, thin curves). (a) Temporal evolution of the phase velocity, computed at $y_0=1$. A jump is observed at $t\approx400$ for SO and at $t\approx 800$ for AI.  (b) Evolution of the velocity profile modulus for SO. The profiles are normalized with respect to their peak value. (c) Evolution of the normalized velocity profiles for the case of AI ic. It can be noticed that the solution for the wake flow admits the presence of near-selfsimilar terms (see also Fig.  \ref{schema}) of expanding velocity profiles. It is also interesting to note that the rapid transition happens when the perturbation reaches the wake from the outer region (as the SO case, panel (b)), or viceversa as  for the AI case (c). The perturbation width follows a power law $t^p$, $p\approx0.42$ in the above cases.}
\label{prof_w}
\end{figure}

The observations of frequency jumps yield an interesting result: the perturbations temporal evolution have a three-part structure, with an early stage, an intermediate stage and an asymptotic stage. This multistage structure  and the nature of the phase velocity jumps discussed in the previous section can be better highlighted by considering  the temporal evolution of the velocity modulus  $|\hat{v}|$ normalized to its instantaneous $\infty$-norm, see Figs. \ref{prof_c} and \ref{prof_w}.

The early transient is the stage where the perturbation is most affected by the fine details of the initial condition and represents a period of adjustment. During the intermediate stage  the perturbation evolves almost exponentially: the phase velocity takes the final constant value, the transverse velocity profiles maintain a near self-similar nature in time and the growth factor changes very slowly. This stage appears as a kind of  \textcolor{black}{temporal intermediate asymptotics} and can be in general considered extinguished only when both the frequency and the temporal growth rate become constant so that the long term regime is reached. This transient structure is common to all perturbations. In the case in which the frequency transient presents one or more jumps, the beginning of the intermediate asymptotics can be envisaged where the near-selfsimilar features first appear. In particular, this first appearance can be observed when the initial conditions  leads to two phase velocity jumps and in between these jumps the phase speed keeps a constant value for many (possibly, up to hundreds or thousands) time scales. See  Figs.  \ref{prof_c} and \ref{prof_w} and the qualitative schematic shown in Fig.  \ref{schema} where the different kinds of phase speed transients that we observed in this work are resumed.
It should be noted that the establishment of self-similarity requires a constant wave propagation speed.

In Figures \ref{prof_c} and \ref{prof_w}, we present  the cases of $Re=100$, $x_0=50$, $k=0.7$, $\phi=\pi/6$ with SO and AI initial conditions for the wake flow and of $Re=6000$, $k=1$, $\phi=\pi/4$ with SC and AW initial condition for the channel flow. In the wake case the similarity is showed by normalizing the solution by the peak value of the profiles  (the $\infty$ - norm) and by normalizing the lateral coordinate by the instantaneous width of the perturbation profile (the distance from the axis where $|\hat{v}|/ \|\hat{v}\|_{\infty}=0.01$). In the channel case,  since the lateral diffusion is blocked by the walls, to highlight the similarity, it was sufficient to normalize the solution by the profile peak value.

\begin{figure}
\begin{center}
\includegraphics[width=\columnwidth]{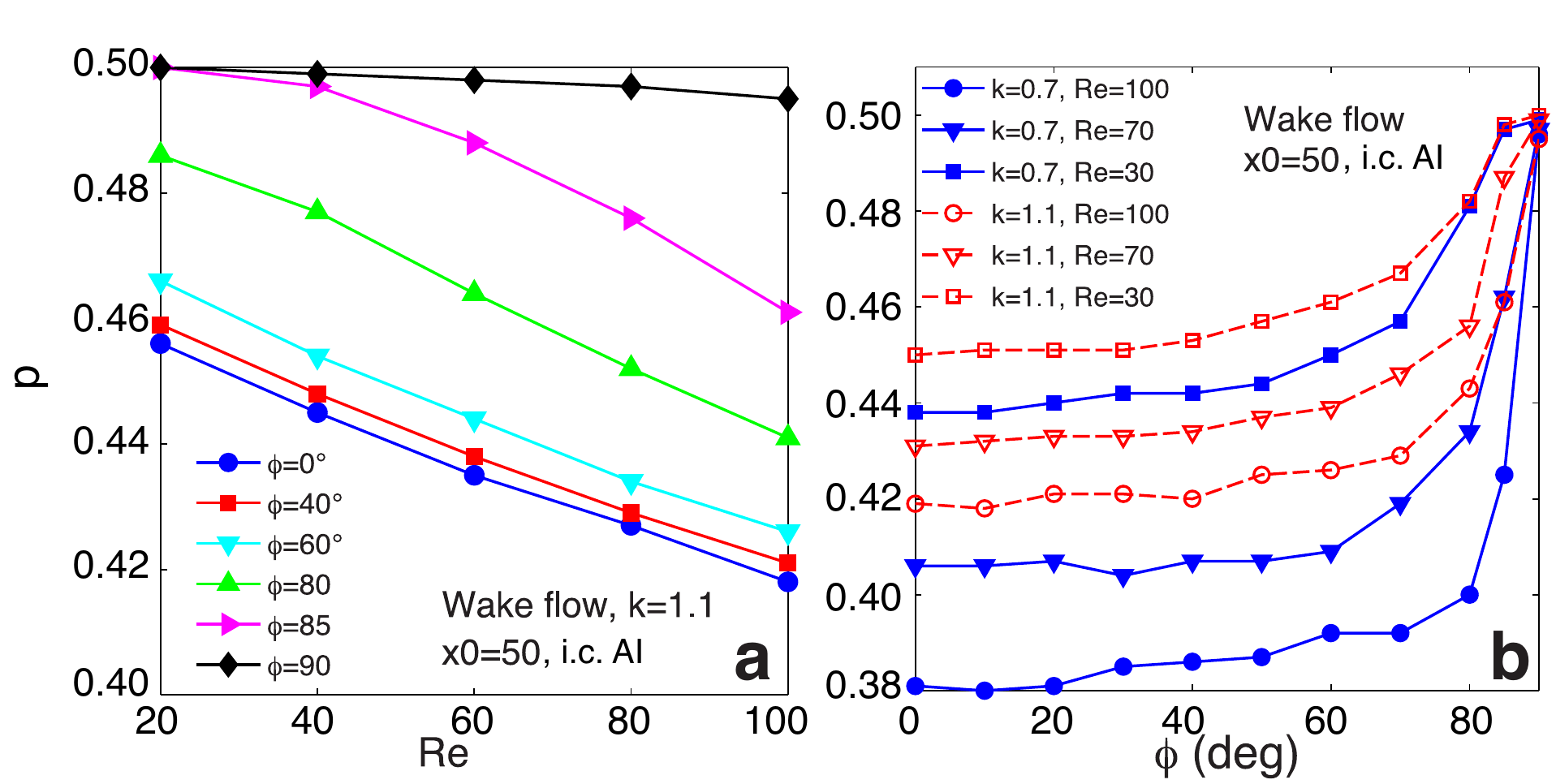}
\caption {Scaling exponent p for the temporal evolution of the wake perturbation width in the intermediate term at $x_0 = 50$ and with AI initial condition. (a) The exponent is represented as a function of the Reynolds number and it is parameterized with the obliquity angle
in the range $[ 0,\pi/2]$. (b) The dependence on the obliquity angle is shown for $Re=\{ 30, 70,100\}$
and $k=\{ 0.7, 1.1\}$.}
\label{exponent}
\end{center}
\end{figure}

\begin{figure*}
\begin{center}
\includegraphics[width=1.99\columnwidth]{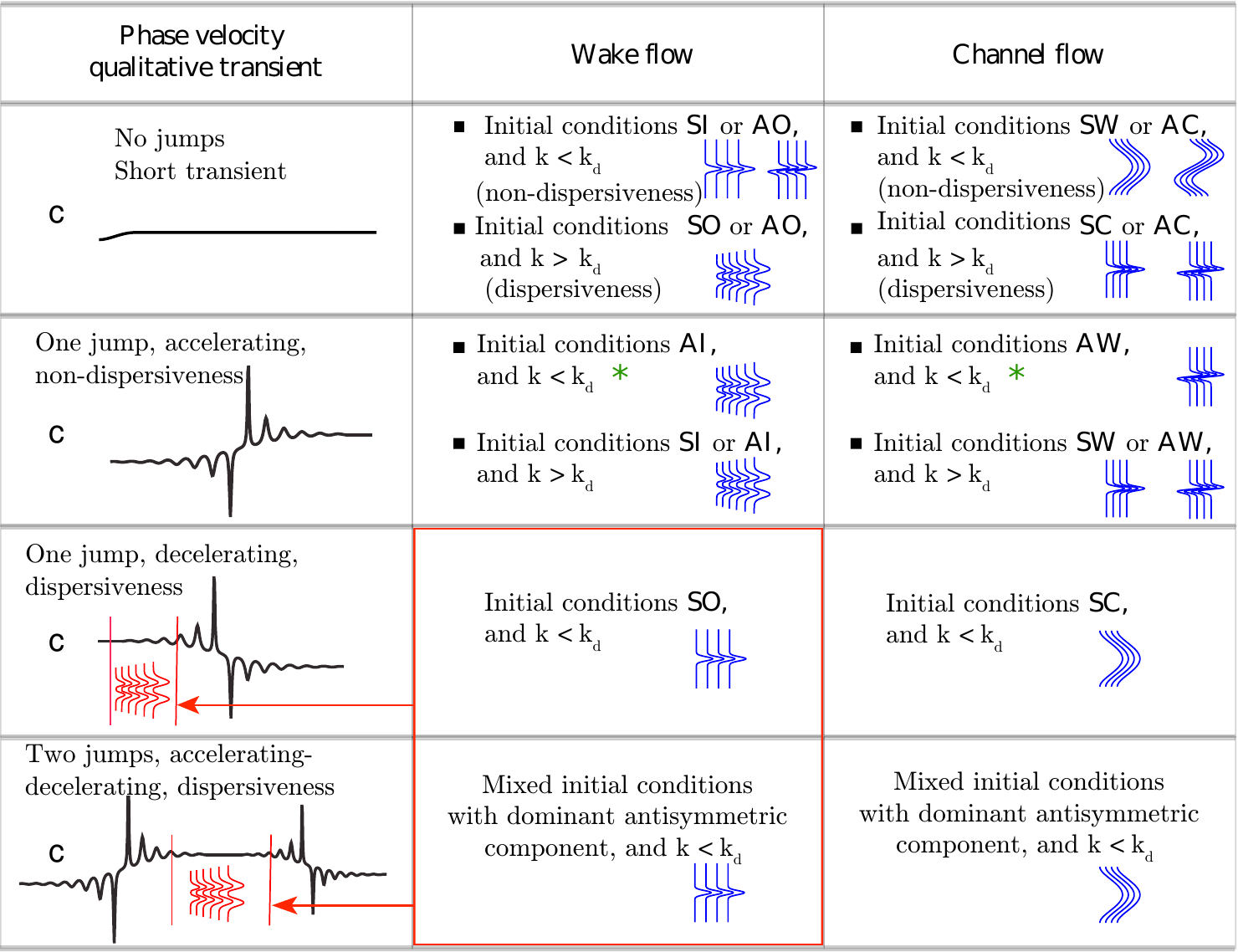}
\end{center}
\caption {\textcolor{black}{Qualitative scheme for the phase velocity transients. \textit{Left column}: type of transient; the black lines represent the phase velocity evolution over time (refer to Figs 4-7 in the paper), while the red sketches indicate the presence of self-similar perturbation profiles, in the intermediate term, for the wake case (please refer to Fig 7). \textit{Mid column}: wake flow; in each cell the conditions under which the corresponding kind of transient realizes are indicated, in terms of initial condition type (SI=symmetric in-wake, AI=antisymmetric in wake, SO=symmetric out-of-wake, AO=antisymmetric out-of-wake) and the wavenumber value, namely $k < k_d$ or $k > k_d$. \textit{Right column}: channel flow. (SC=symmetric central, AC=antisymmetric central, SW=symmetric lateral, AW = antisymmetric lateral). With the blue sketches we represent, for each case, the shape of the perturbation in the long term (temporal asymptotic conditions). The green star is to remind that, even for $k<k_d$, the least stable antisymmetric mode belongs to the right  branch of the spectrum and it is \textcolor{black}{non-dispersive}.} }
\label{schema}
\end{figure*}

In the reported cases, the wake perturbation width scales in time as $t^{p}$, with $p\approx 0.42$. The general trend for $p$ is shown in Fig.  \ref{exponent}.  One can observe that the exponent decreases as the Reynolds number and increases as the wave obliquity and the polar wavenumber.
For very low values of the Reynolds number,  when the inertial effects become very little,  the exponent is expected to take the value 0.5. Note that $Re \approx 20$ is the smallest reliable value for which the wake flow can be represented in terms of a matched asymptotic expansions solution valid in both the spatial intermediate and far field, see Ref. \cite{tb03} for major details on the wake base flow computation. The dependence on the obliquity angle is weak at low angles, while the diffusive scaling $t^{0.5}$ occurs for perturbations orthogonal to the base flow ($\phi=\pi/2$) since in this case the convective transport does not matter, see Eqs.\ref{OS_IVP}-\ref{Squire_IVP}. Moreover, as expected, we observe that short waves present a more diffusive behavior than long ones, see the panel (b) in Fig.  \ref{exponent} .
Fig. \ref{schema} shows a qualitative scheme of the phase velocity transient type, for different combinations of wavenumber -- type of initial condition. The diagram distinguishes the cases where the perturbation is accelerated or decelerated after the jump. For instance, for the wake, in the case of expanding perturbation along the $y$ direction, as shown in Fig.  \ref{prof_w} {\bf b}, the change in the perturbation shape occurs when the phase speed starts to shift to the value specific of the least stable eigenvalue. Other examples of acceleration/retardation of the wave propagation can be find in Figs.  \ref{fig_4_new}, \ref{noise}, \ref{prof_c}, \ref{prof_w}.
It is also noteworthy that the time interval where the frequency/phase velocity jumps are observed is rather distant from the initial moment, and for this reason it could be observed in the laboratory. For instance, in the case of the wake flow with $Re=100$ and AI type of initial condition (see panel {\bf e} in Fig. \ref{fig_4_new}), if we consider a cylinder with $D=2\ cm$, the jump arrives after nearly 4 minutes if the fluid is air, and after nearly 50 minutes if the fluid considered is water. In the case of the channel flow, $Re=6000$ and SC type of initial condition (see panel {\bf a} in Fig. \ref{fig_4_new}), if we consider a channel with $h=15$ cm, the jump arrives after nearly 25 sec if the fluid is air, and after nearly 6 minutes if the fluid considered is water. 

\section{Concluding remarks}

\textcolor{black}{We observed the existence of a wavenumber threshold
$k_d$ that splits the range where the least-damped waves in
the long term  disperse  from the range 
where they are \textcolor{black}{non-dispersive}.
%
%
The transition between dispersive and non-dispersive behavior is related to the fact that for $k < k_d$ 
the least-stable Orr-Sommerfeld eigenvalue belongs to the left branch of the spectrum, while for $k > k_d$, 
it belongs to the right branch. We considered the temporal kinetic energy evolution in the wavenumber space and distinguished the rate of change of kinetic energy due both to viscous  and convection effects. 
It is observed that in the wavenumber space the kinetic energy shows an exchange of mode at $k=k_d$. In fact, the kinetic energy of the solution follows that of the least stable mode which is located on the tip of the left branch of the spectrum for $k < k_d$ and is located on the tip of the right branch for $k \ge k_d$.
The transition in the dispersion relation also helps to explain some features of the transient dynamics. Indeed, initial conditions with a wave number higher or lower than $k_d$ have  different transients. The possible types  of phase velocity transients that we  observed in this study can be classified into four categories:
\begin{enumerate}
\item short transients where the phase velocity reaches its \textcolor{black}{long-term asymptote} smoothly (no jumps)
\item long transients where the phase velocity jumps to a higher value and the wave propagation is accelerated
\item long transient where the phase velocity jumps to a lower value and the wave propagation is decelerated
\item long transients where more than one jump occurs during the phase velocity transient.
\end{enumerate}
%
Jumps are in general accompanied by a modulation of the phase velocity which is observed both before and after the  instant when they take place. This modulation shows a temporal periodicity that is highly correlated with the width of the range of frequencies given by the modal theory.
The type of transient resulting of course depends also on the imposed initial condition. 
Initial conditions can be classified into four types according to their symmetry and to the centrality or laterality of the regions where the initial condition takes its largest variation.
The  observation  of  phase velocity transients leads to identifying  three distinct stages in the time evolution of the perturbations. There is always a first part, the early transient, which is  heavily dependent on the initial condition, a second much longer part, the intermediate transient and a third part, the \textcolor{black}{long-term} state that is reached when both the phase velocity and the growth rate take their final constant values.  Inside the intermediate stage, \textcolor{black}{interesting and non-intuitive}, the perturbation evolves almost exponentially: the phase velocity takes a constant value, the transverse velocity profiles maintain a near self-similar nature over time and the growth factor changes very slowly. In the wake flow, which is a system  slowly evolving in space, our simulations highlight a lateral growth of the perturbation profiles which follows a temporal power law scaling where the exponent changes not only with the Reynolds number, but also with the wavelength and wave-angle.
In the case of purely symmetric or antisymmetric initial condition, the beginning of the intermediate transient is marked by the phase velocity jumps, if there are any.  Transients show two phase speed jumps when the initial condition is antisymmetric and noisy. In this case, near self similarity begins beyond the first jump where the propagation accelerates. }

\section*{Acknowledgments}
The authors thank Ka-Kit Tung, William O. Criminale, Miguel Onorato and Davide
Proment for fruitful discussions on the results presented in this work.

\appendix
\section{A Chandrasekhar eigenfunction expansion method to solve the linearized initial value problem}
In this section, based on an eigenfunctions expansion in terms of the Chandrasekhar functions \cite{chandrasekhar61}, a Galerkin method to solve the Orr-Sommerfeld and Squire initial value problem is presented. The method is valid for every parallel base flow with homogeneous boundary conditions and is conceptually an extension, specific to {\it non modal} calculations, of the use of Chandrasekhar's functions, which satisfy both the vanishing of the normal velocity and its normal derivative on wall boundary conditions of viscous flows. In particular, it is a three dimensional  extension of the method by Gallagher \& Mercer \cite{gm62} which was based on the use of Chandrasekhar's functions to solve the bi-dimensional {\it modal} perturbative problem for the Couette flow.

 As a starting point, the IVP in the wall-normal velocity and vorticity form is considered (this is equivalent to the system \ref{OS_IVP}-\ref{Squire_IVP}).
\begin{flalign}
\label{Orr-Somm}
\dt \dey^2\hv-k^2\dt\hv+i\alpha U(y)\dey^2\hv-i\alpha k^2 U(y)\hv-i\alpha U''(y)\hv\nonumber\\
-\frac{1}{Re}\bigg(\dey^4\hv-2k^2\dey^2\hv+k^4\hv\bigg)=0\\
\label{Squire}
\dt\he+i\alpha U(y)\he-\frac{1}{Re}\bigg(\dey^2\he-k^2\he\bigg)=-i\gamma
U(y)^{\prime}\hv
\end{flalign}
\begin{gather}
\hv(y=\pm 1,t)= \dey\hv (y=\pm 1,t)= \he(y=\pm 1,t)=0\\
\hv(y,t=0)=\hv_0(y)\ \ \ \ \he(y,t=0)={\hat \omega_{y_0}}(y)
\end{gather}

\subsection{Solution to $\hv$ equation}\label{sec:v_solution}

The solution of \eqref{Orr-Somm} can be expressed as a generalized Fourier
expansion with time-dependent coefficients:
\begin{gather}
\label{v_expansion}
\hv(y,t)=\sum_{n=1}^{\infty}c_n(t)X_n(y)\ \ \ \ y \in[-1,1],
\end{gather}
where $X_n(y)$ are orthogonal functions, and the following inverse
transform applies:
\begin{gather}
\label{inverse_trasf}
c_n(t)=\frac{\int_{-1}^1 \hv(y,t)X_n(y)\ud y}{\int_{-1}^1 X_n(y)X_n(y)\ud y}.
\end{gather}
Since in the initial value problem both the initial condition
and the boundary conditions need to be imposed,
it is worthwhile to consider functions that satisfy the boundary conditions.
Moreover note that the coefficients $c_n$ of the series are in general
complex, since $\hv$ is complex-valued and the spatial modes are
considered as real.
The particular orthogonal functions which we use are those defined by
the following fourth order eigenvalue problem
\begin{gather}
\label{problem4}
\frac{\rm d^4}{{\rm d}y^4} X(y)=\lambda^4 X(y)\ \ \ \ \ y \in[-1,1]\\
\label{bc_problem4}
X(y=\pm 1)=0\ \ \ \ \frac{\rm d}{{\rm d}y} X (y=\pm 1)=0.
\end{gather}

\begin{figure}[t]
        \centering
        \advance\leftskip-2.5cm
         \advance\rightskip-2cm
        
	\includegraphics[width=7cm]{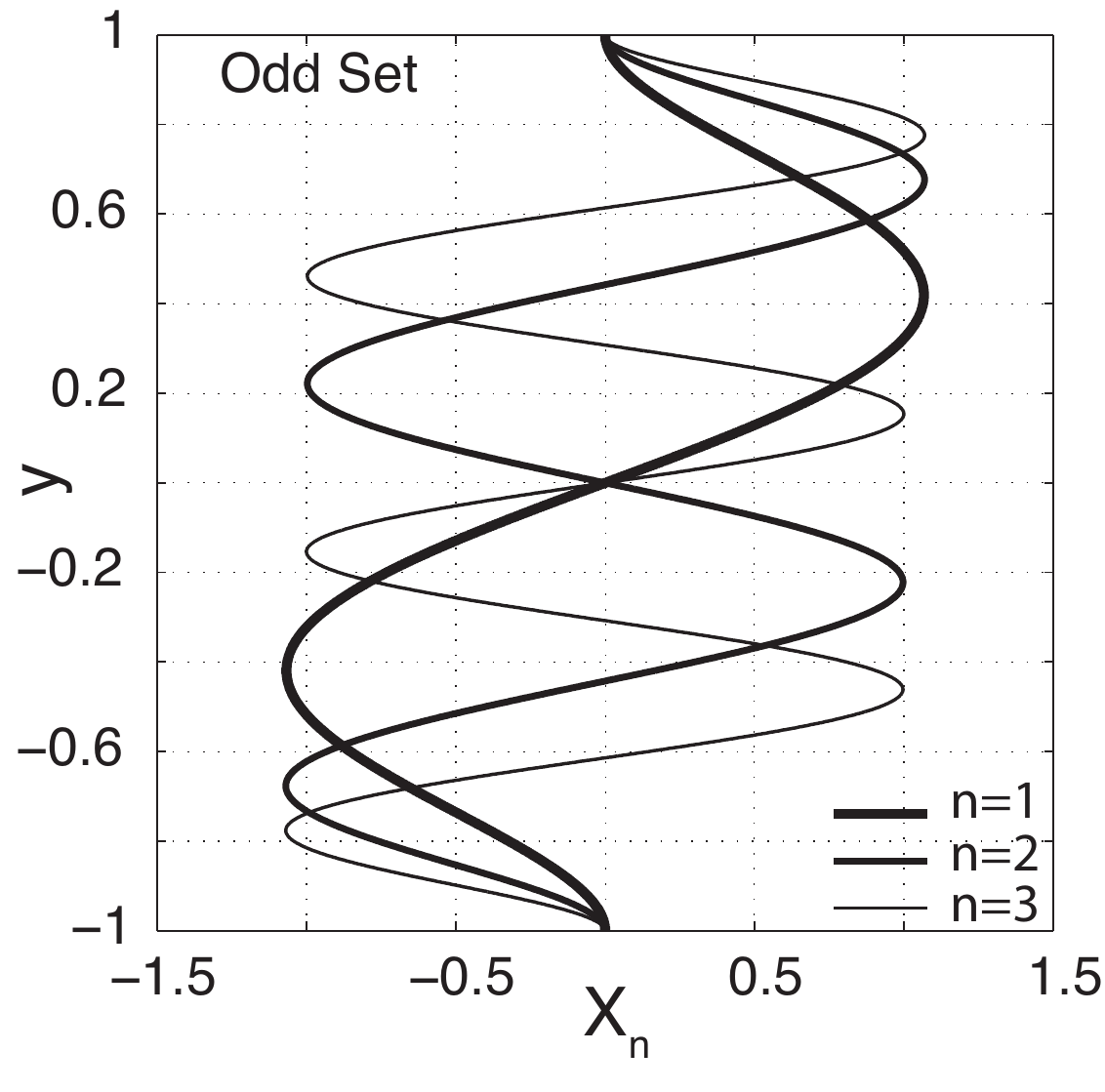}
	\includegraphics[width=7cm]{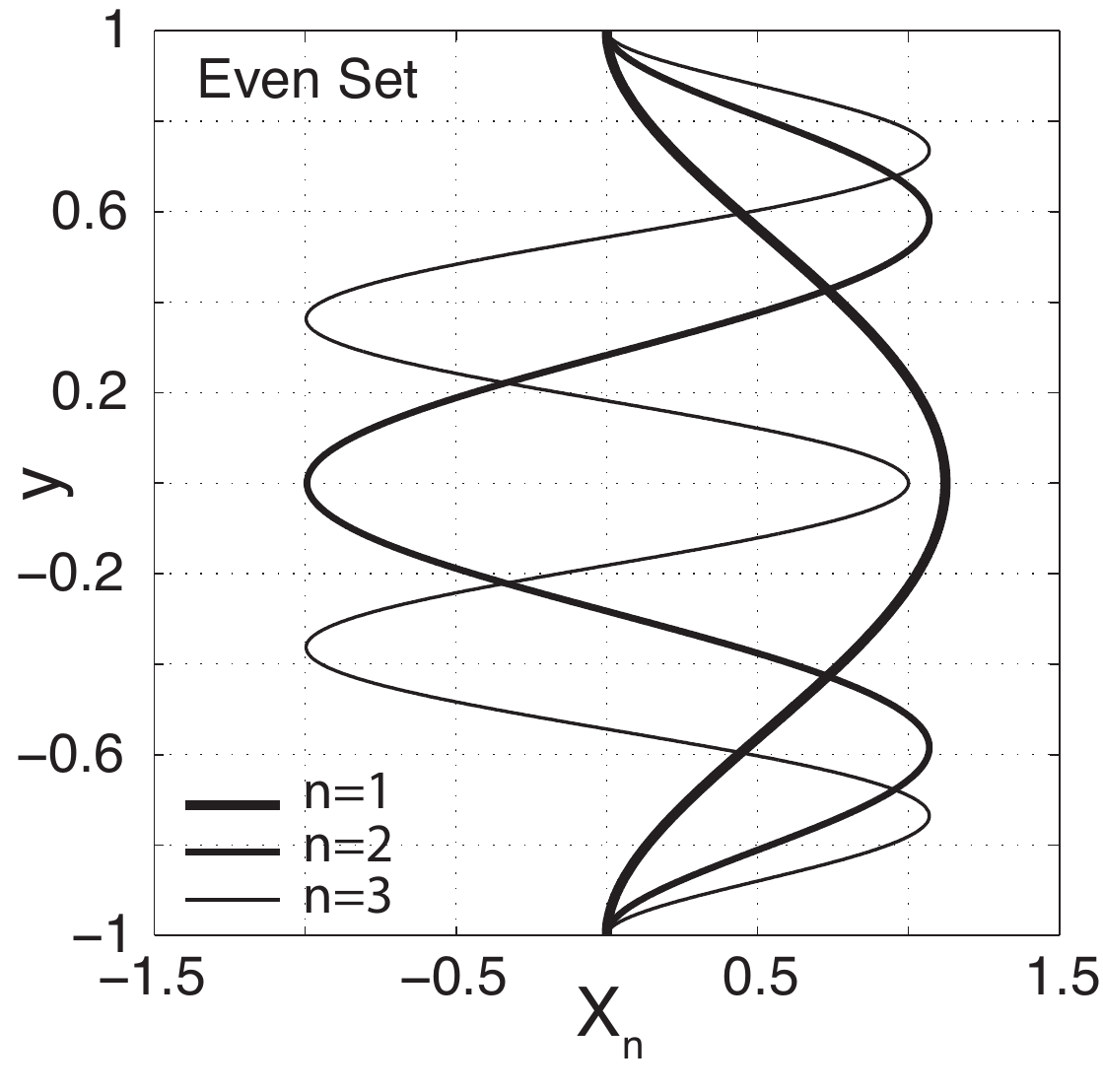}

	\caption{The basis eigenfunctions}
\label{fig:modes_v}
\end{figure}

Two different sets of eigenvalues and the corresponding
eigenfunctions are found, respectively odd and even, by numerically solving
the following transcendental equations
\begin{gather}
\label{trasc_odd}
tan(\lambda_n)-tanh(\lambda_n)=0\ \ \ (odd\
set)\\
\label{trasc_even}
tan(\lambda_n)+tanh(\lambda_n)=0\ \ \ (even\ set).
\end{gather}
The corresponding normalized eigenfunctions
(figure \ref{fig:modes_v}) are
\begin{align}
&X_n=\frac{1}{\sqrt{2}}\bigg[\frac{sinh(\lambda_n
y)}{sinh(\lambda_n)}-\frac{sin(\lambda_n
y)}{sin(\lambda_n)}\bigg]\hspace{0.5cm}{n=1,3,5..,N-1}\nonumber \\ &(odd\
set)\\[10pt]
&X_n=\frac{1}{\sqrt{2}}\bigg[\frac{cosh(\lambda_n
y)}{cosh(\lambda_n)}-\frac{cos(\lambda_n
y)}{cos(\lambda_n)}\bigg]\hspace{0.5cm}{n=2,4,6..,N}\nonumber\\&(even\
set).
\end{align}
Similar functions, in a different domain, have been used in the study of the circular Couette flow between coaxial cylinders \cite[appendix V]{chandrasekhar61}.

Since the imaginary and the real part of the solution $\hv$ usually have
opposite parity, independently on the initial condition, both the odd
and the
even sets are necessary to completely describe the problem and obtain the
correct result. \\
In the following paragraphs a compact notation for the space derivatives is
introduced. In order to simplify the reading, the $y$-derivatives will be
indicated with a subscript. The temporal derivatives will be indicated
explicitly or with a dot. \par
The numerical solution to the $\hv$ equation \eqref{Orr-Somm} is obtained by applying the variational Galerkin method. Truncating the series \eqref{v_expansion} at N functions and substituting yields

\begin{align}
\label{residual}
\epsilon(y,t;\ \alpha,\ \gamma) &=\sum_{n=1}^{N}\dtot{}{t}
c_n(t)X_{n_{yy}}-k^2\sum_{n=1}^{N}\dtot{}{t} c_n(t)X_n\nonumber\\
&+i\alpha U(y)\sum_{n=1}^{N}c_n(t)X_{n_{yy}}-i\alpha k^2 U(y)\sum_{n=1}^{N}c_n(t)X_n\nonumber\\
&-i\alpha\frac{\ud^2 U(y)}{\ud y^2}\sum_{n=1}^{N}c_n(t)X_n-\frac{1}{Re}\sum_{n=1}^{N} c_n(t)X_ {n_{yyyy}}\nonumber\\
&+\frac{2k^2}{Re}\sum_{n=1}^{N}c_n(t)X_{
n_{yy}}-\frac{k^4}{Re}\sum_{n=1}^{N}c_n(t)X_n.
\end{align}

The error functional $\epsilon$ is minimized when it is orthogonal to the space of the linearly independent trial functions $X_n$ with $n={1,2,..N}$. In this context, given two functions $u(y)$ and
$v(y)$ with $y\in\ \Omega=[-1,\ 1]$, the following definition of scalar product applies
\begin{equation}
 \langle u,v\rangle=\int_\Omega u\cdot v\ud y.
\end{equation}
With above notation the Galerkin orthogonality condition is
expressed as
\begin{equation}
 \langle \epsilon,X_m\rangle=0\ \hspace{1.5cm} m=1,2,...,N.
\end{equation}

  The Orr-Sommerfeld PDE is now reduced to a system of $N$
ODEs of the first order, where the time dependent coefficients $c_n(t)$ are the only unknowns.

\begin{align}\hspace{-10pt}
\label{ODE_system}
0 &=\sum_{n=1}^{N}\dtot{}{t}c_n(t)\langle X_{n_{yy}},X_m\rangle-k^2\sum_{n=1}^{N}\dtot{}{t} c_n(t)\langle X_n, X_m\rangle\nonumber\\
&+i\alpha\sum_{n=1}^{N}c_n(t)\langle U(y)X_{n_{yy}},X_m\rangle-i\alpha k^2\sum_{n=1}^{N}c_n(t)\langle U(y)X_n, X_m\rangle\nonumber\\
&-i\alpha \sum_{n=1}^{N}c_n(t)\langle\frac{\ud^2 U(y)}{\ud y^2}X_n X_m\rangle-\frac{1}{Re}\sum_{n=1}^{N}c_n(t)\langle X_{n_{yyyy}},X_m\rangle\nonumber\\
&+\frac{2k^2}{Re}\sum_{n=1}^{N}c_n(t)\langle X_{
n_{yy}},X_m\rangle-\frac{k^4}{Re}\sum_{n=1}^{N}c_n(t)\langle X_n, X_m\rangle
\nonumber\\&n,\ m=1,2,3,...,N.
\end{align}

The scalar products can be
evaluated analytically or computed by numerical integration:\\
\begin{gather}
D_{m,n}=\langle X_n, X_m\rangle=\delta_{m,n}\\[8pt]
S_{m,n}=\langle X_{n_{yy}}, X_m\rangle=\\[8pt] \nonumber
=
\begin{cases}
+4\frac{\lambda_n^2\lambda_m^2}{ \lambda_n^4-\lambda_m^4}
(\lambda_n\mu_n-\lambda_m\mu_m) & \mbox{if } (n+m) \mbox{ is even, }n\ne
m \\
0  & \mbox{if } (n+m) \mbox{ is odd} \\
-\lambda_n^2\mu_n^2+\lambda_m\mu_m & \mbox{if }n=m
\end{cases}\\[15pt]
F_{m,n}=\langle X_{n_{yyyy}}, X_m\rangle=\lambda_n^4\delta_{m,n}\\
U^{(1)}_{m,n}=\langle U(y)X_{n_{yy}},X_m\rangle\\
U^{(2)}_{m,n}=\langle U(y)X_n,X_m\rangle\\
U^{(3)}_{m,n}=\langle\frac{\ud^2 U(y)}{\ud y^2}X_n X_m\rangle,
\end{gather}
where
\begin{equation}
\mu_n=\frac{cosh(2\lambda_n)-cos(2\lambda_n)}{
sinh(2\lambda_n)-sin(2\lambda_n)}\hspace{1cm} \lim_{n\to\infty}\mu_n=1.
\end{equation}

It is convenient to express the ODE system \eqref{ODE_system} in a  compact
notation: in the following, vectors will be indicated either explicitly using
braces or with bold lower case letters; matrices will be indicated with bold
capital letters; constants with roman capital
letters and physical parameters in italic.
The system can be written as
\begin{equation}
\BU{H}\BU{\dot{c}}-\BU{G}\BU{c}=0,\\
\end{equation}
where
\begin{align}
 \BU{H}&=\BU{S}-k^2\BU{D}\\
\BU{G}&=-i\alpha\BU{U^{(1)}}+i\alpha k^2\BU{U^{(2)}}+i\alpha
\BU{U^{(3)}}\nonumber\\
&+\frac{1}{Re}\BU{F}-\frac{2k^2}{Re}\BU{S}+\frac{k^4}{Re}\BU{D},
\end{align}

where $\BU{D}=[D_{m,n}]$ etc., i.e. the element $D_{m,n}$ is placed at the
$n^{th}$ column and at the $m^{th}$ row of the matrix. $\BU{H}$ is invertible,
so denoting $\BU A=\BU{H}^{-1}\BU{G}$ yields
\begin{gather}
\label{ODE_system2}
 \BU{\dot{c}}-\BU{A}\BU{c}=0.
\end{gather}
The complex eigenvalues $\sigma_{i}$ of $\BU{A}$ constitute the spectrum of the Orr-Sommerfeld equation, and the analytic solution to eq. \ref{ODE_system2} can be easily implemented and computed numerically(the QR method is used to compute the eigenvalues, in the matlab environment).
\subsection{Solution to the nonhomogeneous $\he$ equation}\label{sec:eta_solution}

In order to solve the Squire IVP, a set of normal functions different from the one
adopted for the velocity is needed, since the second order PDE only requires $\he$
to vanish at the boundaries, but not its first derivative. A simple
choice for the basis functions, here adopted, is the following
\begin{align}
&Y_n=sin(\xi_n y)\hspace{1.5cm}n=1,3,5,...N-1\ \ &(odd\
set)\\
&Y_n=cos(\xi_n y)\hspace{1.5cm}n=2,4,6,...N\ \ \ &(even\
set),
\end{align}
where
\begin{align}
&\xi_n=\frac{(n+1)\pi}{2}\hspace{1.0cm}n=1,3,5,...N-1\ \ &(odd\
set)\\
&\xi_n=\frac{(n-1)\pi}{2}\hspace{1.0cm}n=2,4,6,...N\ \ \ &(even\
set).
\end{align}
Also in this case, note that two sets of eigenfunctions are put together to
form a unique set, since both are necessary to completely describe the
complex-valued normal vorticity.
The general solution is then obtained as the sum of a particular solution
${\hat \omega_{y_p}}$ and the solution to the corresponding homogeneous equation ${\hat \omega_{y_h}}$
\begin{equation}
 \he(y,t)={\hat \omega_{y_{h}}}(y,t)+{\hat \omega_{y_{p}}}(y,t)
\end{equation}
\begin{equation}
 \he(y,t)=\sum_{n=1}^{\infty}(b_{h_{n}}+b_{p_{n}})(t)Y_n(y).
\end{equation}
Applying the Galerkin method yields to the following forced ODE set

\begin{gather}
\BU{\dot{b}}-\underbrace{\big(\
-i\alpha\BU{U^*}+\frac{1}{Re}\BU{S^*}-\frac{2k^2}{Re}\BU{D^*}\big)}_{
\BU{G^*} } \BU { b }=\underbrace{-i\gamma\BU{F^*}}_{\BU{B}}\BU{c}\\
\BU{\dot{b}}-\BU{G^*}\BU{b}=\BU{B}\BU{c},
\label{Squire_ODE_sys}
\end{gather}

where the eigenvalues of $\BU{G^*}$ constitute the spectrum of the Squire equation and
\begin{gather}
 D^*_{m,n}=\langle Y_n,Y_m\rangle=\delta_{m,n}\\
 S^*_{m,n}=\langle Y_{n_{yy}},Y_m\rangle=-\xi_n^2\delta_{m,n}\\
 U^*_{m,n}=\langle U(y)Y_n,Y_m\rangle\\
 F^*_{m,n}=\langle\dtot{U(y)}{y}X_n,Y_m\rangle.
\end{gather}

The homogeneous solution $\BU{b_{h}}$ of \eqref{Squire_ODE_sys} is analytically known, while a particular solution $\BU{b_{p}}$ of the following form is sought
\begin{equation}
\label{b_p}
b_{p_n}(t)=\sum_{j=1}^N a_{nj}e^{\sigma_jt},
\end{equation}
where $a_{nj}$ are constants and $\sigma_j$ are the eigenvalues of $\BU{G}$, in such a way the forced solution has the same spectral content of forcing term. The coefficients are obtained through the solution of $N$ algebraic sets, after the computation of $\BU{c}(0)$.\par
\begin{figure}
        \centering
	\includegraphics[width=7.7cm]{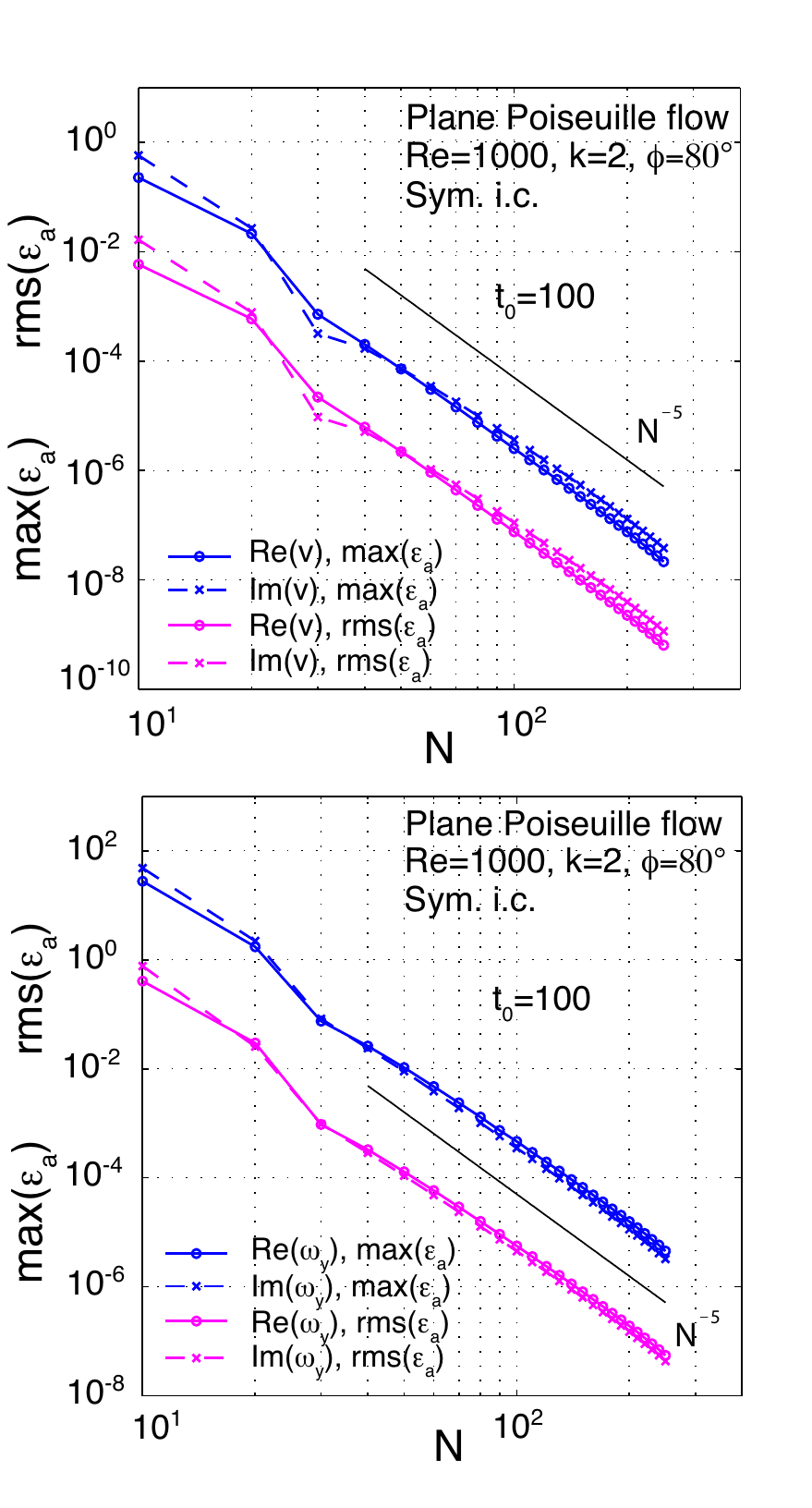}
	\caption{Maximum and rms of the absolute error of $\hv$ (left panel) and $\he$ (right panel) as a function of the number of modes $N$ for channel flow with $t_0=100$, $Re=1000$, $k=2$, $\phi=80^\circ$ and symmetrical initial
condition. Continuous line: real part. Dashed line: imaginary part. Magenta line: maximum absolute error. Blue line: rms of the absolute error. Black line: accuracy trend $N^{-5}$ \cite{orszag1971}. Since the exact solution is not known, the residuals are defined as the difference between the solution and an accurate solution computed with 350 modes.   $\epsilon_a(y,t)=|\hv_N(y,t)-\hv_{N=350}(y,t)|$,\hspace{1cm}$rms(\epsilon_a)(t)=\frac{1}{N_y}\sqrt{\sum_{i=1}^{N_y}\epsilon_a^2(y,t)}$,\hspace{1cm}$max(\epsilon_a)(t)=\max_{y_i}(\epsilon_a(y,t))$}
\label{fig:Convergence}
\end{figure}

The Galerkin method was first applied to the Orr-Sommerfeld modal equation by \cite{dolph1958}. They used normal functions that guarantee a $1/N^4$ convergence ratio. Gallagher \cite{gm62} used, for the modal problem, the Chandrasekhar-Reid functions and the error decreased as $1/N^{5}$ with $N\to\infty$ as shown in \cite{orszag1971}. The fifth order of accuracy is ensured for the present formulation as well, as shown in figure \ref{fig:Convergence}.

The method results to be fast and accurate in time and space.  Since the time evolution is analytically represented, the complete wave transient, up to the asymptote, can be simulated without typical drawbacks of time-marching techniques. Arbitrary initial conditions can be specified for bounded flows. The limits of the method are related to the non-normality of the Orr-Sommerfeld and Squire operators. The non-normal effects act on the numerical procedure by worsening the condition number of the eigenvector matrices. Anyway, the sensibility of the spectrum (especially at high $Re$ and $k$ values) is a property of the stability operator and it is thus independent on the numerical scheme. 

\section{Spectra computation}
\begin{figure}
        \centering
	\includegraphics[width=7cm]{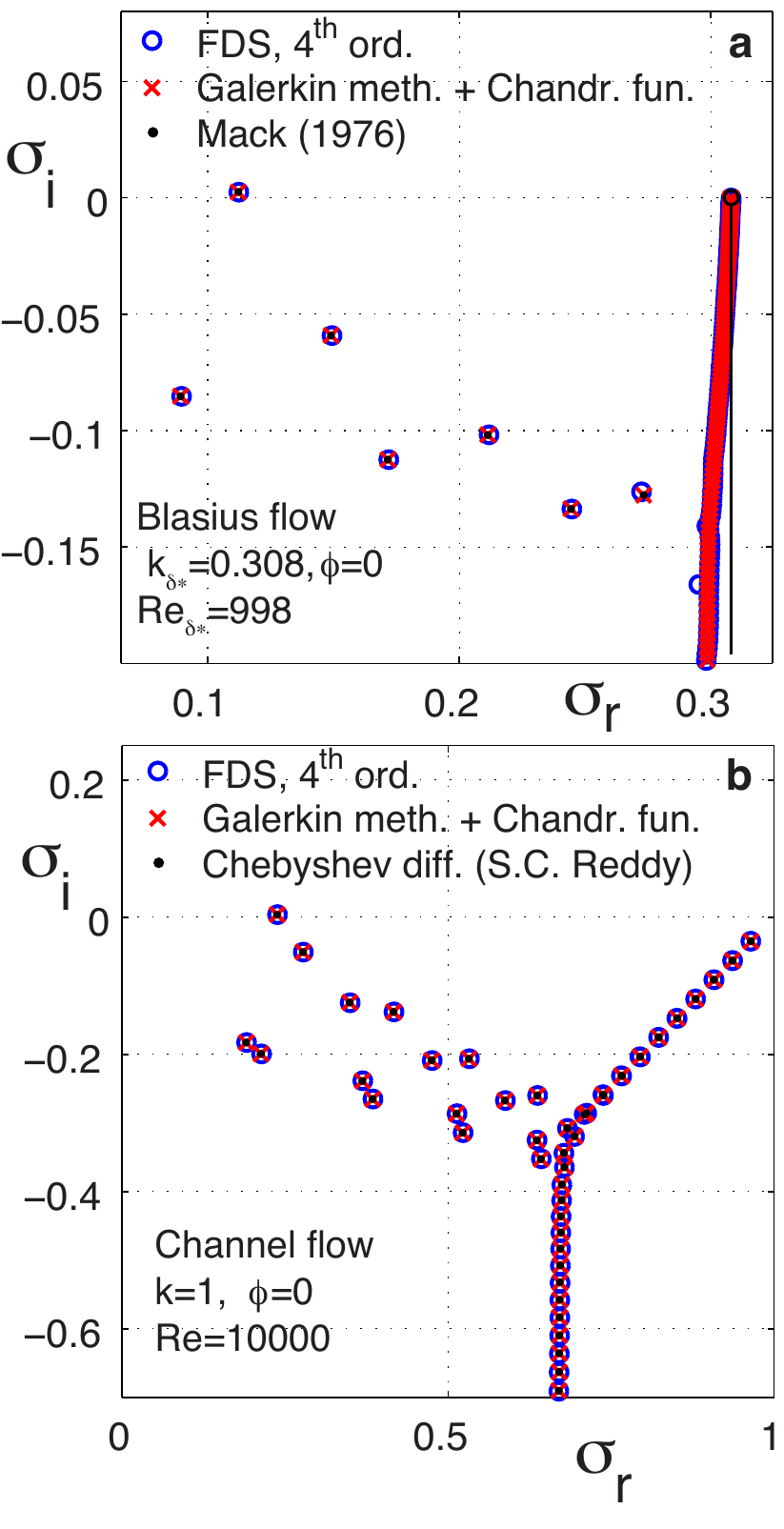}
	 \caption{Eigenvalues spectra ($\sigma=\sigma_r+i\sigma_i$) of the Orr-Sommerfeld equation. Comparison of different numerical methods: a $4^{th}$ order finite differences scheme on uniform grid (blue circles); a $5^{th}$ order Galerkin method on nonuniform grid (red x);  a Chebyshev spectral collocation method (code by S.C. Reddy, black points on panel {\bf (b)}); results by Mack \cite{m1976} (black points on panel {\bf a}). {\bf(a)} Blasius boundary layer flow, $Re_{\delta^{*}}=998$, $k_{\delta^{*}}=0.308$, $\phi=0$.  The continuous part of the spectrum is discretely approximated. The black line represents the analytic solution, obtainable only if the boundary condition at $y\to\infty$ is relaxed to $\hat{v}_{y\to\infty} bounded$. As $Re_{\delta^{*}}$ and $k_{\delta^{*}}$ increase, particular attention is needed to avoid spurious eigenvalues due to the spectrum intrinsic sensibility. {\bf (b)} Channel flow, $Re=10000$, $k=1$, $\phi=0$.}
\label{fig:spectra_comparison}
\end{figure}
Two different numerical methods are used in the present work to compute the spectra of the Orr-Sommerfeld equation: - the fourth order finite differences collocation scheme \cite{ascher1994} and - the present non modal three dimensional version of the Gallagher and Mercer (1962) method which is described here in Appendix A. Here, we report a comparison with literature results as a validation of our spectral calculations, see Fig.12. For unbounded flows it has been shown by Grosch \cite{gs1978} that a continuous spectrum can be analytically found, if the boundary conditions are relaxed to $\hat{v}\ bounded$ as $y\to\infty$. If finite-norm  boundary conditions are imposed and therefore only the class of decaying solutions as $ |y|\to\infty$ is considered, the continuous part of the spectrum is approximated in a discrete way. If the boundary conditions are imposed far from the wake, the approximation is very good. The Galerkin method with Chandrasekhar functions described above was successfully adapted to the wake flow and to the boundary layer flow. Since no spectra with our wake base flow (see \cite{tb03}) have been found in literature, the schemes have been validated with the Blasius boundary layer flow (see figure  \ref{fig:spectra_comparison}, panel a). Eventually, for the channel flow, the comparison with a hybrid spectral collocation method based on Chebyshev polynomials \cite[A.6]{sh2001} is shown in figure \ref{fig:spectra_comparison}, panel b.


\bibliographystyle{apsrev}
\bibliography{PRE_bibliography_new}

\end{document}